\documentclass[
reprint,
nofootinbib,
amsmath, 
amssymb,
aps,
]{revtex4-2}

\usepackage{ulem}
\usepackage{graphicx}
\usepackage{dcolumn}
\usepackage{bm}
\usepackage{tabularx}

\usepackage{orcidlink}

\newcommand\ba{\begin{eqnarray}}
\newcommand\ea{\end{eqnarray}}
\newcommand{\sv}{\langle \sigma v\rangle}

\newcommand{\eq}[1]{\begin{alignat}{2} #1 \end{alignat}}

\newcommand{\cor}[1]{{\textcolor{red}{#1}}}
\newcommand{\cog}[1]{{\textcolor{green}{#1}}}

\newcommand{\cms}{{\rm cm}^3/{\rm s}~}
\newcommand{\ee}{e^+e^-}
\newcommand{\gamgam}{\gamma\gamma}
\newcommand{\DM}{{\rm DM}}

\newcommand{\mnras}{MNRAS}
\newcommand{\apjl}{The Astrophysical Journal Letters}
\newcommand{\apjs}{The Astrophysical Journal Supplement Series}
\usepackage{subfig, caption}
\DeclareCaptionJustification{justified}{\leftskip=0pt \rightskip=0pt \parfillskip=0pt plus 1fil}
\usepackage{hyperref}
\hypersetup{
	colorlinks=true,
	linkcolor=blue,
	filecolor=magneta,      
	urlcolor=blue,
}
\begin{document}

\preprint{APS/123-QED}

\title[DM annihilation in EoR]{  
Machine Learning Constraints on Dark Matter Annihilation during the Epoch of Reionization: A CNN Analysis of the 21-cm Signal
}
%
%
%
\author{Atsushi J. Nishizawa\,\orcidlink{0000-0002-6109-2397}}
\email{atsushi.nisizawa@gifu.shotoku.ac.jp}
\affiliation{Gifu Shotoku Gakuen University,
Takakuwanishi, Yanaizu, Gifu, 501-6194, Gifu, Japan \\
Institute for Advanced Research, and Kobayashi Maskawa Institute, Nagoya University,
Furosho, Chikusaku, Nagoya, 464-8602, Aichi, Japan}
\author{Pravin Kumar Natwariya\,\orcidlink{0000-0001-9072-8430}}
\email{pvn.sps@gmail.com}
\affiliation{School of Fundamental Physics and Mathematical Sciences, Hangzhou
Institute for Advanced Study, UCAS, Hangzhou, 310 024, China}
\affiliation{University of Chinese Academy of Sciences, Beijing, 100 190, China}
\author{Kenji Kadota\,\orcidlink{0000-0003-2019-6007}}
\email{kadota@ucas.ac.cn}
\affiliation{School of Fundamental Physics and Mathematical Sciences, Hangzhou
Institute for Advanced Study, UCAS, Hangzhou, 310 024, China}
\affiliation{University of Chinese Academy of Sciences, Beijing, 100 190, China}
%
%
%
%
%
%
\date{\today}

\begin{abstract}
We explore the impact of dark matter annihilation on the 21-cm signal during the cosmic dawn and epoch of reionization (EoR). Using modified 21cmFAST simulations and convolutional neural networks (CNNs), we investigate how energy injected into the intergalactic medium (IGM) through dark matter annihilation affects the evolution of the 21-cm differential brightness temperature. Focusing on two annihilation channels, photon-photon ($\gamgam$) and electron-positron ($\ee$), we examine a broad range of dark matter masses and annihilation cross-sections. Our results show that CNNs outperform traditional power spectrum analysis by effectively distinguishing between subtle differences in simulated 21-cm maps produced by annihilation and non-annihilation scenarios. We also demonstrate that the structure formation boost, driven by dark matter clumping into halos and subhalos, significantly enhances the annihilation signal and alters the thermal and ionization history of the IGM. This enhancement leads to a noticeable effect on the 21-cm signal, including a shift from absorption to emission as dark matter annihilation heats the IGM at lower redshifts.
By incorporating observational noise from upcoming radio interferometers, particularly the Square Kilometer Array (SKA), we show that these effects remain detectable despite observational challenges. 
We find that the dark matter annihilation models can leave measurable imprints on the 21-cm signal distinguishable from the non-annihilation scenarios for the dark matter masses $m_\DM=100$ MeV and the annihilation cross-sections of $\sv \simeq 10^{-31}~\cms (\sv \simeq 10^{-32}~\cms$ for $m_{\DM}=1$ MeV and $\sv \simeq 10^{-24}~\cms$ for $m_{\DM}=1$ TeV).

\end{abstract}

\maketitle

\section{Introduction}
\label{sec:intro}

A redshifted 21-cm signal--- due to the hyperfine transition between 1S singlet and triplet states of the neutral hydrogen atom, is expected to provide unprecedented insight into the period when the first luminous objects formed. Dark matter plays a crucial role in shaping the evolution of cosmic dawn and reionization. Despite the searches for decades, the nature of dark matter remains one of the most profound mysteries in modern cosmology, and particle physics \cite{Planck:2018, Boddy:2022knd, Cooley:2022ufh, Feng:2010}. 
In this paper, we explore the annihilation of the dark matter and subsequent effects on the 21-cm signal. In particular, we explore the potential of machine learning techniques, specifically convolutional neural networks (CNNs), to differentiate between annihilating dark matter models and the conventional $\Lambda$CDM without dark matter annihilation using simulated three-dimensional 21-cm differential brightness temperature maps from the cosmic dawn era. We also explore the dark matter annihilation boost due to structure formation, which can significantly impact the thermal and ionization history of the Universe--- thus 21-cm signal at lower redshifts ($\lesssim 30$) \cite{Liu:2016, Liu_2020, Amico:2018}. To generate realistic 21-cm maps, we employ a modified version of the publicly available {\tt 21cmFAST} code \cite{Mesinger:2011FS}. This semi-numerical simulation allows us to efficiently explore a range of dark matter masses and annihilation cross-sections. We produce brightness temperature maps on $200^3$ grids with a box size of 300 Mpc/$h$, providing sufficient resolution to capture relevant spatial fluctuations while maintaining computational feasibility.

Previous analyses of the 21-cm signal have predominantly focused on the summary statistics with an emphasis on the power spectrum. Although these methods have been effective, they may not capture all the information encoded in the full three-dimensional brightness temperature field. This limitation becomes particularly significant when trying to distinguish subtle differences between various dark matter models. Machine learning, as a rapidly advancing field, offers a substantial potential for uncovering complex patterns within high-dimensional data \cite{Murakami:2020, Sabiu:2022, Murakami:2024, Murakami::2024}. In the context of 21-cm cosmology, machine learning techniques, particularly deep learning methods such as CNNs, offer a novel approach to data analysis \cite{Gillet:2019, Villanueva:2021, Sabiu:2022, Murakami::2024}. These methods can identify features in the 21-cm maps that can not easily captured by traditional statistics, enhancing our ability to discriminate between different dark matter models and analyze observational data.

The application of CNNs to 21-cm data analysis builds upon recent successes in other areas of astrophysics and cosmology. For instance, CNNs have been used effectively in galaxy classification \cite{Zhu:2019, Cheng:2021, Wu:2023}, gravitational wave detection \cite{Gebhard:2019, Li:2020, Xia:2021, Baltus:2021, Zhang:2022}, and CMB foreground removal \cite{Farsian:2020, Casas:2022, Yan:2024}. In the context of dark matter studies, machine learning has already shown promise in improving limits from gamma-ray observations and in analyzing direct detection experiments \cite{Caron:2018, Khek:2022, Khosa:2020}. 
For less massive dark matter, the large-scale structure is smoothed out on small scales due to the dark matter free streaming, which is imprinted in the neutral hydrogen gas distribution at the post-reionization epoch and can be effectively constrained by the CNN \citep{Murakami::2024}.
CNNs can automatically learn hierarchical features from the data. In the context of 21-cm differential brightness temperature maps, these networks can identify complex spatial patterns that arise due to dark matter annihilation and subsequent effects on structure formation and astrophysical processes during the cosmic dawn. This approach can be a treasure trove for detecting the subtle effects of dark matter models that produce only small deviations from the non-annihilation scenario.

\section{Simulations and Initial Conditions}
\label{sec:simulation}

\subsection{21-cm Differential Brightness Temperature}

This subsection focuses on the 21-cm differential brightness temperature, $\Delta T_b$--- which is the observable quantity, as a tracer of the dark matter annihilation. We consider the epochs before and at the start of reionization. The epochs are observed in the form of absorption/emission by the neutral hydrogen medium relative to the cosmic microwave background radiation ($T_{\rm CMB}$) or background radio radiation ($T_{\rm R}$) at a reference wavelength of 21-cm. The corresponding frequency for the 21-cm line (hyperfine transition between 1S singlet and triplet states of the neutral hydrogen atom) is $\nu_0=1420.4$~MHz. To explore the IGM state at a redshift of $z$ using the 21-cm signal observations, the corresponding present-day observed frequency can be mapped as $1420.4/(1+z)$~MHz. The 21-cm differential brightness temperature \cite{Pritchard_2012, Natwariya:2023T}, 
\eq{
\Delta T_b=\frac{T_{\rm exc}-T_{\rm R}}{1+z}\ (1-e^{-\tau_{\nu_0}})\,,
}
here $T_{\rm R}$ is the background radio radiation temperature. In the present study, we consider the background radiation solely by CMB, i.e. $T_{\rm R}=T_{\rm CMB}$. For the neutral hydrogen gas and the 21-cm radiation, the excitation temperature, $T_{\rm exc}$, is the spin temperature, $T_{\rm S}$. The spin temperature of the neutral hydrogen atoms is defined by the number density ratio in the hyperfine states--- 1S singlet and triplet states.  In the absence of any external sources, the hyperfine transition probability per atom is once in $\sim 10^7$ years. The presence of any exotic source of energy injection into the medium can significantly affect the hyperfine transition. In the cosmological scenarios, there are mainly three processes that can affect the evolution of the spin temperature: (I) background radio radiation, (II) Ly-$\alpha$ radiation from the first stars or exotic sources--- for e.g. decaying/annihilating dark matter particles, (III) collisions of a hydrogen atom with other hydrogen atoms and residual free protons or electrons,
\eq{
T_{\rm S}^{-1}=\frac{ T_{\rm R}^{-1}+x_\alpha\, T_\alpha^{-1}+x_c\, T_{\rm gas}^{-1} }{1+x_\alpha+x_c}\,,
}
here $T_\alpha$ is the colour temperature of Ly-$\alpha$ radiation from first luminous objects and dark matter annihilation. $x_c$ is collisional coupling coefficient, and $x_\alpha$ is the Ly-$\alpha$ coupling coefficient due to Wouthuysen-Field effect \cite{Pritchard_2012, Natwariya:2023T, 1952AJ.....57R..31W, Field, 1958PIRE...46..240F, 2006MNRAS.367..259H, 2011MNRAS.411..955M},
\eq{
x_{c} = \frac{T_{\nu_0}}{T_R}\frac{C_{10} }{A_{10} }\ , \ \  x_{\alpha} = \frac{T_{\nu_0}}{T_R}\frac{P_{01} }{A_{10} }\ , \label{xcxa}
}
here, $C_{10}=N_ik_{10}^{iH}$ is collision deexcitation rate and $T_{\nu_0}=2\,\pi\,\nu_0=0.068~{\rm K\,}$. $i$ stands for hydrogen atom, electron and proton. $k_{10}^{iH}$ is the spin deexcitation specific rate coefficient due to collisions of species $i$ with hydrogen atom \cite{2012RPPh...75h6901P}. $P_{01}=4P_{\alpha}/27$, and $P_\alpha$ is scattering rate of Ly-$\alpha$ radiation \cite{2012RPPh...75h6901P}. $A_{10}=2.85\times 10^{-15}$~sec$^{-1}$ is the Einstein coefficient for triplet to singlet state spontaneous emission in neutral hydrogen. The Wouthuysen-Field (WF) coupling coefficient can be written as \cite{2012RPPh...75h6901P, 2011MNRAS.411..955M},
\begin{alignat}{2}
x_\alpha=1.7\times10^{11}\,(1+z)^{-1}\, S_\alpha\,J_\alpha\,,
\end{alignat}
here, $S_\alpha=\int dx\, \phi_\alpha(x)\,J_\nu(x)/J_\infty$, is of order unity and involves atomic physics. It describes the photon distribution near Ly-$\alpha$ resonance. $J_\infty$ is the flux far from the absorption. $J_\alpha$ is the specific Ly-$\alpha$ photon flux. Furthermore, $T_{\rm gas}$ refers to the gas temperature. The optical depth of the HI medium can be given by \cite{Pritchard_2012, Natwariya:2023T, Mesinger:2011FS},
\eq{ 
\tau_{\nu_0} \simeq \frac{3\ n_{\rm HI}}{32\, \pi\, \nu_0^3}\ \frac{T_{\nu_0}}{T_{\rm S}}\ \frac{A_{10}}{H(z)}\ \frac{H(z)/(1+z)}{\partial v/\partial r}\,,
}
here, $n_{\rm HI}$ represents the neutral hydrogen number density and $H(z)$ represents the Hubble parameter. Furthermore, $\partial v/\partial r$ is the proper velocity gradient of HI along the line of sight, and it can be taken as $H(z)/(1+z)$ for high redshift or in the absence of peculiar velocity. In the above equation, term $\big([H(z)/(1+z)]/[\partial v/\partial r]\big)$ can be written as $\big(H(z)/[\partial v_r/\partial r+H(z)]\big)$ in the comoving coordinates; where $\partial v_r/\partial r$ is the comoving gradient of comoving velocity of HI along the line of sight component \cite{Mesinger:2011FS}. The optical depth of the HI is $\tau_{\nu_0}\ll1$, and therefore, the 21-cm differential brightness temperature can be written as,
\eq{
\Delta T_b\simeq  &27\,x_{\rm HI}\,(1+\delta_b)\left(1-\frac{T_{\rm R}}{T_{\rm S}}\right) \left(\frac{\Omega_{\rm b}\,h^2}{0.023}\right) \nonumber\\
&\times \sqrt{\frac{0.15}{\Omega_{\rm m }\,h^2}\ \frac{1+z}{10}}\ \left(\frac{H(z)/(1+z)}{\partial v/\partial r}\right) ~{\rm mK},
\label{eq:DeltaTb}
}
here $x_{\rm HI}$ is neutral hydrogen fraction and $\delta_b$ is the baryon energy density contrast. Hereafter, we will refer to the 21-cm differential brightness temperature simply as the 21-cm signal for convenience. 

The 21-cm signal dimensionless power spectrum can be defined as,
\eq{
\Delta_{21}^2(k,\,z)= \frac{k^3}{2\,\pi^2\, V}\ \langle\Delta T_b(\bm k,\,z)\Delta T_b^*(\bm k,\,z)\rangle_k\,,\label{eq:powersp}
}
where $V$ is the volume of the box and $\Delta T_b(\bm k,\,z)$ is the Fourier transformation of $\Delta T_b(\bm x,\,z)$. To get the global 21-cm signal or the average 21-cm signal ($\Bar{\Delta T_b}(z)$) of the box at redshift $z$, we can divide the addition of $\Delta T_b(\bm x,\,z)$ in every cell by the number of total cells at a corresponding redshift.

In the next subsection, we discuss how dark matter annihilation affects the evolution of the 21-cm signal. We use the C-only version of the publically available code {\tt 21cmFAST v2.1}\footnote{\href{https://homepage.sns.it/mesinger/DexM\_\_\_21cmFAST.html}{https://homepage.sns.it/mesinger/DexM\_\_\_21cmFAST.html}} with some required modifications to include the dark matter annihilation in the evolution of the 21-cm signal maps \cite{Mesinger:2011FS}.

\subsection{Sources of Energy Injection}
\label{ssec:energyinjection}
The annihilation of dark matter can inject energy into plasma via energetic particles. The energy injection per unit time and per unit volume,
\begin{alignat}{2}
	\frac{dE}{dV\,dt}\bigg|_{\rm inj} = \,\rho_{\rm0,\DM}^2\ f_{\rm DM}^2\,(1+z)^6\,\frac{\langle \sigma\,v\rangle}{m_{\rm DM}}\,,\label{eq:inj}
\end{alignat}
here $\rho_{\rm 0,\DM}$ is present-day dark matter energy density and $f_{\DM}$ represents the fraction of dark matter which annihilate. Furthermore,  $\sv$ counts for the thermal average of the cross-section and relative velocity of dark matter particles and $m_{\DM}$ represents the mass of the dark matter. The plasma may not absorb the particles produced by dark matter annihilation at the time of their production.  Different species can have different mean free paths depending on their energy at the time of production and on the state of plasma. Therefore, deposited energy into plasma can be different for different species,
\begin{alignat}{2}
    \frac{dE}{dV\,dt}\bigg|_{{\rm dep},c} = f_c(z,\,m_{\DM})\ \frac{dE}{dV\,dt}\bigg|_{\rm inj}\,, \label{eq:dep}
\end{alignat}
here function, $f_c(z,\,m_{\DM})$, represents the energy deposition efficiency into the plasma by dark matter annihilation for corresponding channel $c$ \cite{Liu:2016}, 
\eq{
f_c(z,\,m_{\rm DM})=&\frac{H(z)}{(1+z)^{\alpha-3}}\ \int  dz'\  \frac{(1+z')^{\alpha-4}}{H(z')}\,\nonumber\\
&~~~~~\times T_c(z',\,z,\,m_{\rm DM})\ [1+\mathcal{B}(z')]\,,\label{fcbz}
}
here, $\alpha=6$ for $s$-wave annihilation (and 8 for a velocity dependent $p$-wave annihilation cross section etc). In the present work, for simplicity, we consider the velocity-independent cross section ($s$-wave self-annihilation scenarios). $T_c(z',\,z,\,m_{\DM})$ is the transfer function, and it accounts for the energy fraction deposited into the plasma at redshift $z$ and injected by dark matter annihilation at redshift $z'$ \cite{Liu:2016}. The transfer function is different for every produced species. We, for illustration, consider two cases where dark matter annihilates into: $\ee$ or $\gamgam$. There are three channels ($c$) to deposit energy into the plasma: (i) $f_{\rm heat}(z,\,m_{\DM})$, heating of the medium (ii) $f_{\rm exc}(z,\,m_{\DM})$, Ly-$\alpha$ excitation of the Hydrogen atoms, and (iii) $f_{\rm ion}(z,\,m_{\DM})$, ionization of Hydrogen and Helium. 
We also allow for the enhancement of the dark matter annihilation due to the structure formation compared to the uniform dark matter background density, represented by the so-called boost factor $(1+\mathcal{B})$, for which we adopted the parameterization in the DarkHistory package \cite{Liu:2016, Liu_2020}. For the case of instantaneous deposition, $f_c(z,\,m_{\rm DM})=1/3 \times(1+\mathcal{B}(z))$, i.e., an equal amount of energy goes to heating, excitation and ionization of the medium.
We go beyond such a simple ``on-the-spot" approximation for the energy deposition efficiency and adopt a more involved treatment using {\tt DarkHistory}\footnote{\href{https://darkhistory.readthedocs.io/en/master/}{https://darkhistory.readthedocs.io/en/master/}} package \cite{Liu_2020}. {\tt DarkHistory} differs from a simple on-the-spot approximation, and it can compute the energy deposition efficiency into the plasma at every step of redshift as a function of dark matter mass and the annihilation channel. We computed the energy deposition efficiency, ionization fraction and gas temperature for different dark matter parameters (the parameters are $m_{\rm DM},\sv$) at $z=30$ by {\tt DarkHistory}, which was then used as the initial conditions for {\tt 21cmFAST} code.

We consider homogenous evolution of IGM temperature and ionization fraction before redshift 30--- as the formation of luminous structures comes into effect after $z\sim30$ \cite{Pritchard_2012}. In the presence of energy deposition into the plasma by dark matter annihilation, the gas temperature evolution with redshift \cite{1999ApJ...523L...1S, 2000ApJS..128..407S, 2019JHEP...02..187C, 2018PhRvL.121a1103D, 2009PhRvD..80b3505G}, 
{\small\eq{
    \frac{dT_{\rm gas}}{dz}  =  \frac{2\,T_{\rm gas}}{(1+z)} + \frac{\Gamma_{C}}{(1+z)\,H}\, (T_{\rm gas}-T_{\rm CMB})+\frac{dT_{\rm gas}}{dz}\bigg|_{\DM}\,.
	\label{dtdz} 
}}
The last term in the above equation stands for dark matter annihilation,
{\small\eq{
 \frac{dT_{\rm gas}}{dz}\bigg|_{\DM} = - \frac{1}{(1+z)\,H}\ \frac{2}{3\,n_{\rm H}\,(1+f_{\rm He}+x_e)}\ \frac{dE}{dV\,dt}\bigg|_{\rm dep,heat}\,,
}}
here, $f_{\rm He}=n_{\rm He}/n_{\rm H}$ stands for the Helium fraction. The  Compton scattering rate is defined as,
\begin{equation}
	\Gamma_{C}= \frac{8\, \sigma_T\, a_r T_{\rm CMB}^4\, x_e}{3\,(1+f_{\rm He}+x_e)\,m_e}\,,
\end{equation}
here, $\sigma_T$, $a_r$ and $m_e$ represent the cross-section for Thomson scattering, Stefan-Boltzmann radiation constant and mass of the electron, respectively. In the above equation, $x_e=n_e/n_{\rm H}$ represents the ionization fraction. Where $n_e$ and $n_{\rm H}$ are the free electron and hydrogen number densities, respectively.  The evolution of ionization fraction with redshift in the presence of energy injection by dark matter annihilation \cite{1999ApJ...523L...1S, 2000ApJS..128..407S, 2018PhRvL.121a1103D, 2011PhRvD..83d3513A, 2009PhRvD..80b3505G},
{\small\eq{
    \frac{dx_e}{dz} &= \frac{\mathcal{P}}{H\,(1+z)}\times\,
    \Big[ \,n_{\rm H}\, x_e\, x_{\rm HII}\,\alpha_{\rm H} \nonumber \\
    &\hspace{1.5em}- 4\,(1-x_{\rm HII})\,\beta_{\rm H} \,e^{-E_{\alpha}/T_{\rm CMB}} \Big]+\frac{dx_e}{dz}\bigg|_{\rm DM}\,.
	\label{dxedz}
}}
The contribution to the ionization of the IGM due to the dark matter annihilation,
{\small\eq{
\frac{dx_e}{dz}\bigg|_{\rm DM} &=  - \frac{1}{H\,(1+z)} \ \frac{1}{n_{\rm H}}\ \Bigg[\,\frac{1}{E_0}\, \frac{dE}{dV\,dt}\bigg|_{\rm dep,ion} \nonumber \\
        &\hspace{8.5em}+\frac{1-\mathcal{P}}{E_\alpha}\, \frac{dE}{dV\,dt}\bigg|_{\rm dep,exc}\,\Bigg]\,,
}}
here $x_{\rm HII}$ is the free proton fraction and can be taken as $x_{\rm HII}\equiv x_e$ in the absence of helium.  $\alpha_{\rm H}$ and $\beta_{\rm H}$ are the case-B recombination coefficient and photo-ionization rate, respectively \cite{1999ApJ...523L...1S, 2000ApJS..128..407S}. Furthermore, $E_0=13.6$~eV and $E_\alpha=(3/4)\, E_0$ are ground state binding energy and Ly-$\alpha$ transition energy for the hydrogen atom, respectively. $\mathcal{P}$ is the Peebles coefficient--- decay probability of a hydrogen atom from the $n=2$ state to the ground state before photoionization occurs \cite{Peebles:1968ja, AliHaimoud:2010dx, 2018PhRvL.121a1103D},
\begin{alignat}{2}
    \mathcal{P}=\frac{1+K_{\rm H}\,\Lambda_{\rm H}\,n_{\rm H}\,(1-x_{\rm HII})}{1+K_{\rm H}\,(\Lambda_{\rm H}+\beta_{\rm H})\,n_{\rm H}\,(1-x_{\rm HII})\,}\,,
	\label{}
\end{alignat}
here, $K_{\rm H}=\pi^2/(E_\alpha^3\, H)$ and $\Lambda_{\rm H}=8.22/{\rm sec}$ account for redshifting of Ly-$\alpha$ photons due to expansion of the Universe and the 2S-1S level two photon decay rate of hydrogen atom, respectively \cite{1984PRA...30....1175P}.

After the redshift of $\sim30$, the spatial distribution of baryons begins to play a significant role in the IGM evolution as the formation of luminous structures starts to take place. Therefore, in order to obtain the thermal and ionization history of the IGM, it is necessary to keep track of the local evolution of gas. The thermal and ionization  equations, (\ref{dtdz} \& \ref {dxedz}), are modified as \cite{Mesinger:2011FS},
{\small\eq{
\frac{dT_{\rm gas}(\bm x,\,z)}{dz} &= \frac{2\,T_{\rm gas}}{3\,n_b}\ \frac{dn_b}{dz}-\frac{T_{\rm gas}}{1+x_e}\ \frac{dx_e}{dz} \nonumber\\ 
&-\frac{1}{(1+z)\,H} \ \frac{2}{3\,(1+x_e)}\sum_p \epsilon_p +\frac{dT_{\rm gas}}{dz}\bigg|_{\rm DM}\,,
}}
{\small\eq{
\frac{dx_e(\bm x,\,z)}{dz} &= \frac{1}{(1+z)\,H}\ \big[\, \alpha_A\,C\,x_e^2\,n_b\,f_{\rm H}-\Lambda_{\rm ion}\, \big] + \frac{dx_e}{dz}\bigg|_{\DM}\,,
}}
here $n_b\equiv n_b(\bm x,\,z)$ is the hydrogen and helium number density. $\epsilon_p\equiv \epsilon_p(\bm x,\,z)$ stands for the Compton and X-ray heating of the gas. Furthermore, $\alpha_A$ represents the case-A recombination coefficient; $C=\langle n_e^2\rangle/\langle n_e\rangle^2$ represents the clumping factor for the free-electrons on the scale of the simulation cell; $f_{\rm H}=n_{\rm H}/(n_{\rm H}+n_{\rm He})\,$; and $\Lambda_{\rm ion}$ accounts for the ionization rate per baryon.

For the case of dark matter annihilation, we obtain three main contributions for Ly-$\alpha$ flux: (i) excitation of neutral hydrogen by X-ray, (ii) stellar emission between  Ly-$\alpha$ and Lyman limit, (iii)  Ly-$\alpha$ flux due to collisional excitations of neutral hydrogen resulting from energy injection in plasma by dark matter annihilation. The contribution from dark matter annihilation in Ly-$\alpha$ flux can be written as,
\begin{alignat}{2}
J_{\alpha,\,\rm DM}= \frac{1}{8\pi^2\nu_\alpha^2\,H^2}\ \frac{dE}{dV\,dt}\bigg|_{\rm dep,exc}\,,
\end{alignat} 
here, $\nu_\alpha$ and $H$ stands for Ly-$\alpha$ frequency and Hubble parameter, respectively. $dE/dV/dt|_{\rm dep,exc}$ represents the energy deposition rate per unit volume in the plasma by dark matter annihilation via the excitation channel.

\section{Machine Learning Method}
\label{sec:ML}

In this section, we describe the image-based discrimination model using Convolutional Neural Network and the method of training and performance evaluation procedures.

\subsection{CNN model}
\label{ssec:CNN}
In this paper, we use the standard simple convolutional neural network architecture, which is widely used in the literature.
The input data can be either 2-dimensional pixels, where the image is defined on the $N\times N$ regular grid, or 3-dimensional voxels, where the image is defined on the $N\times N\times N$ regular grid. In either case, we 
start with the single-channel input. 
{The CNN model consists of 4 convolutional layers, where each layer involves the following sequence of operations: convolution with a kernel size of 3, activation using the Leaky ReLU function, max pooling with a kernel size of 2 and stride of 2, and finally, batch normalization.}

The first convolution maps the single channel to 32 and subsequently increases the number of channels to 64, 128 and 256. Before transferring to the fully connected layer, we put a dropout layer with a 50\% rejection probability. The final output is a single continuous value between 0 and 1, produced by the sigmoid function. If the output is 1, the input is predicted as the annihilation model; an output of 0 indicates it is predicted as the non-annihilation model.

We explore the possible CNN architectures which effectively extract the information from the image without losing information content as much as possible. The simplest case is that the input is only a batch sample of the image. We find that the 21-cm brightness temperature maps of the annihilation and non-annihilation scenarios look fairly similar. More specifically, the fluctuation patterns show little difference, and the most prominent difference is in the global amplitude and scatter. This means that after subtracting the mean from the image and dividing it by the standard deviation, the images from the two models look almost identical. Such a transformation likely happens during batch normalization. Therefore, we first take the mean and standard deviation from the original image and concatenate them to the latent vector before transferring it to the fully connected layer. We find that in the case of a small training set, the additional information of mean and standard deviation slightly helps improve the model discrimination, but once the number of training samples is large enough, the improvement is negligible.

\subsection{Fine Tuning and Training}
\label{ssec:training}
The most ambiguous part of the machine learning-based method is optimizing the architecture for the given problem and for the given dataset available. There are several ways to systematically optimize the hyperparameters, including grid search, Latin Hypercube search, Bayesian optimization, or even the method based on the genetic algorithm \cite{GA1992}. However, in this paper, we do not systematically explore the most optimal configuration of the architecture but instead find out the combination of hyperparameters in an empirical manner. This hyperparameter set is not necessarily the most optimal, but we demonstrate that it works for our simulation set, and it is subject to be optimized in the practical application.

We prepare 1,000 realizations for the conventional model without dark matter annihilation and also for every dark matter annihilation model, where each realization has a box of 300 Mpc/$h$ on a side, with a voxel resolution of 1.5 Mpc/$h$. The size of the voxel corresponds to the angular size of $4.6, 5.1$ and $5.9$ arcmin at redshift $z=10, 13$ and $16$, respectively (it corresponds to the line of sight resolution, in terms of a frequency resolution, of 150 kHz at 1.4 GHz).
The 300 Mpc/$h$ simulated box is divided into $4\times 4\times 4$ subregions where each 3-D image has $50\times 50\times 50$ voxels. In total, we have 128,000 images: half of them are for the no-annihilation scenarios, and the rest are for the annihilation scenarios. 
We randomly draw 20\% of the data as the hold-out sample (test set), while the remaining 80\% is reserved for the training sample. We further split this 80\% into a training set and a validation set. The training set is used for optimizing the model, while the validation set is used to evaluate the model's performance and monitor for overfitting. Importantly, the validation set is not used for model optimization. We stop the training when the validation loss no longer improves, indicating that further training might lead to overfitting.

\section{Results and Discussion}
\label{sec:result}

We begin by presenting our findings for dark matter annihilation scenarios with a dark matter mass of $m_\DM = 10^2$ MeV and a cross-section of $\sv = 10^{-28}~\cms$ through the electron-positron annihilation channel ($\ee$) as a fiducial model.

We first focus on the dark matter mass of $m_\DM = 10^2$ MeV because this mass leads to the most efficient energy absorption into the gas for the redshift range of interest. The cross-section $\sv = 10^{-28}~\cms$ is selected as an example because this value corresponds to the current upper bounds from X-ray and CMB while still being large enough to produce visually noticeable effects on the 21-cm signals for illustration purposes \cite{Cirelli:2023tnx,2016PhRvD..93b3527S}. 
Therefore, we consider the dark matter annihilation model with the mass of $m_\DM=10^2$~MeV and the cross-section of $\sv=10^{-28} ~\cms$ through the electron-positron channel ($\ee$) as a fiducial model. The other parameter sets will also be discussed in the subsequent sections. 

\subsection{Map generation}
\label{ssec:map_generation}
\begin{figure*}
    \begin{center}
        \includegraphics[width=\linewidth]{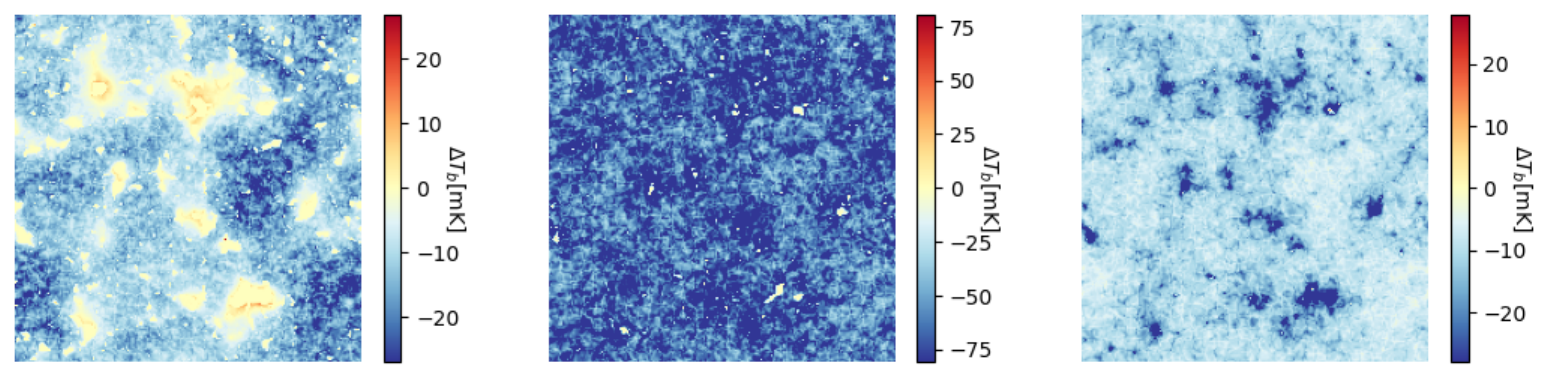}
        \includegraphics[width=\linewidth]{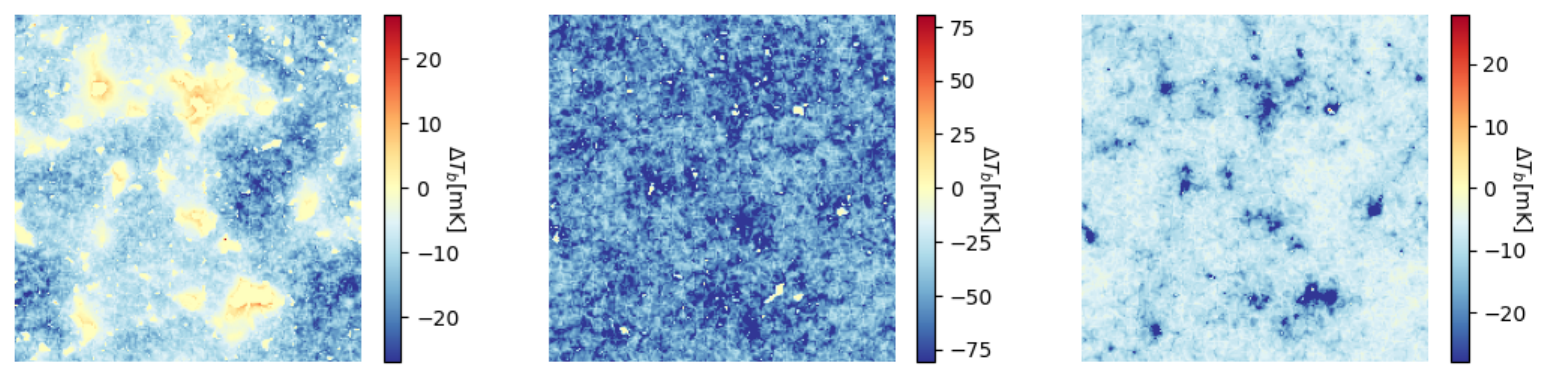}
        \includegraphics[width=\linewidth]{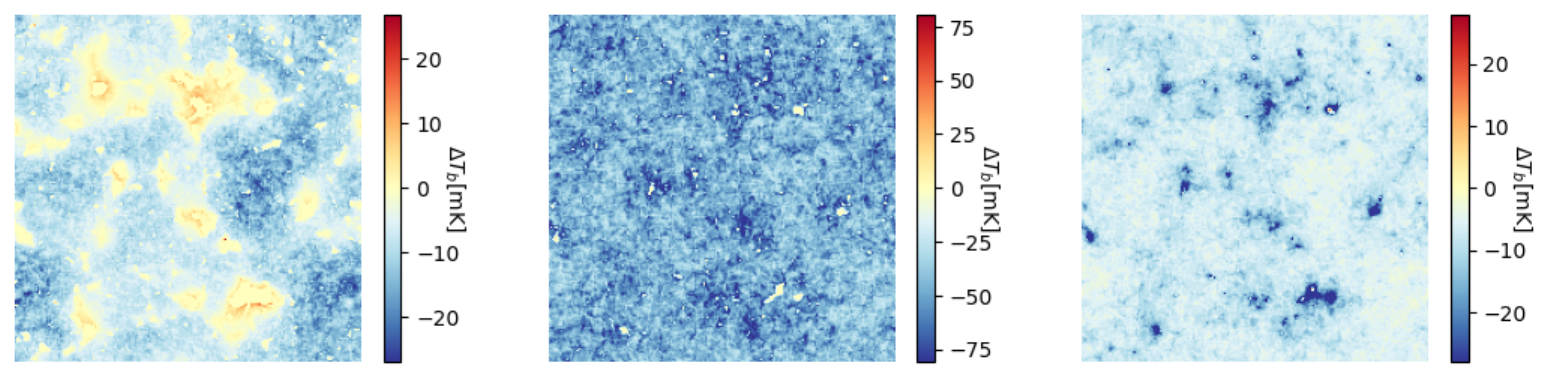}
    \end{center}
    \caption{21-cm differential brightness temperature snaps at redshifts 10, 13 and 16 from left to right. The upper panels represent the conventional $\Lambda$CDM model without dark matter annihilation, and the middle panels are for the annihilation scenarios with the dark matter particle mass $m_{\rm DM}= 100$~MeV, and cross-section $\sv=10^{-28}~\cms$ for $\gamma\gamma$ channel. The lower panels are the same as the middle but for the $\ee$ channel.
    \label{fig:maps}}
\end{figure*}
\begin{figure*}
    \begin{center}
        \subfloat[] {\includegraphics[width=3.2in,height=2in]{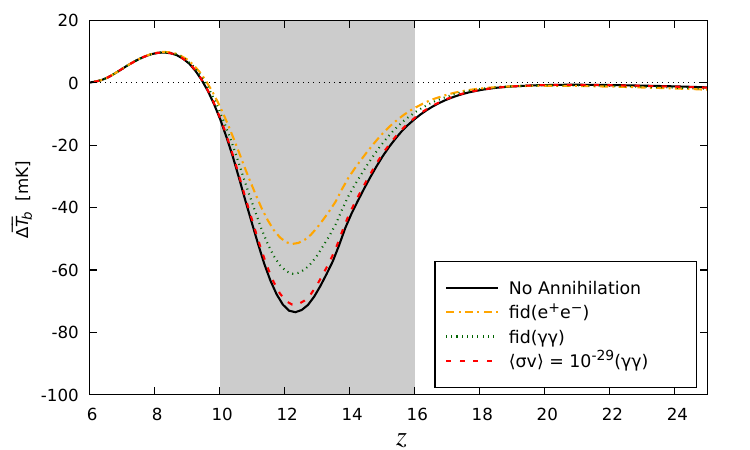}\label{fig:21cm:a}}
        \subfloat[] {\qquad\includegraphics[width=3.2in,height=2in]{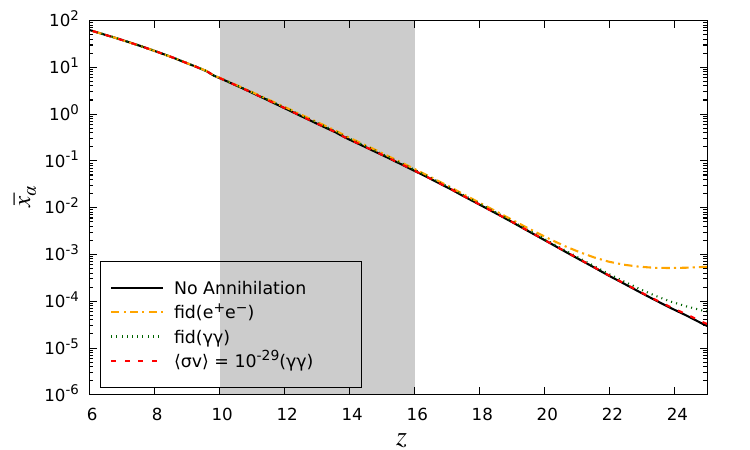}\label{fig:21cm:b}}  
    \end{center}
    \caption{
    Left: Global 21-cm signal as a function of redshift. Right:  Global Ly-$\alpha$ coupling coefficient as a function of redshift. In the present work, we consider the shaded region to explore the effect of the dark matter annihilation on IGM evolution--- particularly redshifts 10, 13 and 16. The dotted black horizontal line is added for reference in the left figure, which represents the scenario with no 21-cm signal, i.e. no absorption and no emission ($\Delta T_b=0$). In both figures, No Annihilation refers to the conventional $\Lambda$CDM scenario without dark matter annihilation. Furtheremore, fid() represts the annihilation case for dark matter particle mass $m_{\rm DM}=100$~MeV, and cross-section $\sv=10^{-28}$~cm$^3$/s for corresponding channels. The red dashed line represents the case with dark matter annihilation into $\gamma\gamma$ channel for $m_{\rm DM}=100$~MeV, and $\sv=10^{-29}~\cms$.}\label{fig:21cm}
\end{figure*}

The brightness temperature maps are generated on the $200^3$ regular grid across the box size of 300 Mpc/$h$. Figure \ref{fig:maps} shows the 21-cm brightness temperature map for different dark matter models at redshift $z=10, 13$ and 16.
The top panel represents the case with no dark matter annihilation, while the middle and lower panels represent the annihilation of dark matter into $\gamma\gamma$ and $e^+e^-$ channels, respectively, on the scale of 300~Mpc. It can be seen here, that as we go from top to bottom panels, the darker blue regions decrease--- showing a lower absorption amplitude.  According to the standard scenario of the star formation, the first luminous objects start to form around redshift $z\sim 25$. The Ly-$\alpha$ photons from the first stars start to couple the spin temperature to the gas temperature around redshift $z\sim 20$, and we can see an absorption profile in the 21-cm line as shown in the right panels of Figure \ref{fig:maps}.

In Figure \ref{fig:21cm}, we plot the global 21-cm differential brightness temperature and global Ly-$\alpha$ coupling as a function of redshift. In the figure, the black solid line represents the scenario without dark matter annihilation.  Considering the annihilation of dark matter decreases the amplitude of the 21-cm signal as the heating of the IGM gas rises--- shown by the coloured lines in \ref{fig:21cm:a}. The case with $\sv=10^{-29}$~cm$^3$/s and $m_{\rm DM}=100$~MeV is shown by the red dashed line for dark matter annihilating to photons $\DM \DM \rightarrow \gamma \gamma$. As we increase the annihilation cross-section to $10^{-28}$~cm$^3$/s, the IGM gas temperature rises, resulting in a lower amplitude of 21-cm signal--- shown by green dotted and yellow dot-dashed lines. For $\DM \DM \rightarrow \ee$ (yellow dot-dashed line), we obtain a higher gas temperature compared to the $\gamgam$ channel (green dotted line). 
{
It happens because the mean free path of the electrons/positrons is smaller compared to the photons for considered dark matter mass \cite{Evoli:2012, Slatyer:2009}. The larger fraction of energy injection for the emitted electrons/positrons happens through the inverse-Compton (IC) process when the energy of $\ee$ is above 1~MeV. 
For the photons, the energy injection happens through pair production and Compton scattering. During the considered redshift range, we remain well within the Thomson limit--- i.e. the energy of the incoming photon in the rest frame of the electron remains very small compared to the energy of the electron \cite{Valds:2010}. For example, the CMB photons have energy about $3\times10^{-3}$~eV at a redshift of 13, and the injected electrons have energy of $10^8$~eV. This sets the upper and lower limits on the upscattered CMB photon energy between $E_{\rm in}^\gamma$ and $4\gamma_e^2E_{\rm in}^\gamma$, where $E_{\rm in}^\gamma$ is the energy of the initial photon, and $\gamma_e$ is the Lorentz factor of the electron. Therefore, at a redshift of 13, an electron having the energy of 100~MeV can upscatter CMB photons in the energy range from $\sim0.003$~eV to $\sim0.5$~KeV, which can ionize the atoms. While electrons having the energy of 1~GeV can upscatter CMB photons to an energy of $>10$~KeV, which can free stream into the plasma until their energy is redshifted to $<10$~KeV. Next, the electrons having an energy of 10~MeV can upscatter CMB photons to an energy of about 5~eV, which is insufficient for the ionization. Therefore, the peak of energy deposition efficiency appears around 100~MeV for all three considered energy deposition channels \cite{Liu:2016}. 
}
Therefore, the $\ee$ channel loses energy faster compared to the $\gamma\gamma$ channel--- it results in a lower 21-cm signal amplitude for the $\ee$ channel. The same has been shown in Figure \ref{fig:maps}.
As we go towards lower redshift, the gas temperature continuously decreases due to the expansion of the Universe, resulting in a higher absorptional signal (Figure \ref{fig:21cm:a}). Around a redshift $z=14$, the radiation from the first stars starts to dominate over the adiabatic cooling of the IGM. This results in a higher gas temperature, but still, Ly-$\alpha$ coupling remains below unity. Around a redshift of $z \sim 12.5$, $x_\alpha$ becomes larger than unity (see Figure \ref{fig:21cm:b}), and we can see a turning point in Figure \ref{fig:21cm:a}--- as the IGM gas temperature already started rising. We obtainthe maximum absorption amplitude at about this redshift--- also shown by middle panels of Figure \ref{fig:maps}. As we go towards a lower redshift, the gas temperature continuously rises. In addition to this, the neutral hydrogen fraction ($x_{\rm HI}$) also decreases due to the ionizing radiation from the first stars. Therefore, we obtain a lower absorptional amplitude for 21-cm signal at redshift $z\sim 10$ as $T_{21}\propto x_{\rm HI}(1-T_{\rm CMB}/T_{\rm S})$ (left panels of Figure \ref{fig:maps} and Figure \ref{fig:21cm:a}). In Figure \ref{fig:maps}, the white regions correspond to $T_{21}\sim 0$, i.e. either no neutral hydrogen or $T_{\rm S}\sim T_{\rm CMB}$. At a later time, around a redshift of 10, the Ly-$\alpha$ coupling becomes $x_\alpha\gg1$, leaving $T_{\rm S}\sim T_{\rm gas}$. 
Additionally, some patches are significantly ionized and become nearly transparent for 21-cm radio photons, resulting in a no-signal--- left panels of Figure \ref{fig:maps}. In contrast, at higher redshifts, no-signal regions correspond to $T_{\rm S}\sim T_{\rm CMB}$ because there are not enough Ly-$\alpha$ photons to couple the gas and spin temperature--- right panels of Figure \ref{fig:maps}. After a redshift of about 9, X-ray heating dominates over dark matter annihilation heating and all lines get merged--- Figure \ref{fig:21cm:a}. Therefore, we consider three values of redshift (10, 13 and 16) to explore the dark matter annihilation effects on IGM evolution and constrain the dark matter parameter space. The Ly-$\alpha$ coupling due to the photons from dark matter annihilation is important at higher redshifts At lower redshifts, the photons from the first stars dominate Ly-$\alpha$ coupling. 

In Figure \ref{fig:21cm:b}, the black solid line represents the Ly-$\alpha$ coupling due to the first stars. The coloured lines also include the effects of dark matter annihilation on the  Ly-$\alpha$ coupling.  During the dark ages, there are no first stars hence no Ly-$\alpha$ photons from the first stars. The coupling between gas and spin temperature is determined by the collisions between hydrogen atoms and other gas species--- $x_c$ in equation \eqref{xcxa}. As the Ly-$\alpha$ coupling due to dark matter annihilation also contributes to the coupling between spin and gas temperature during the Dark Ages, it can modify the evolution of the 21-cm signal during the Dark Ages. In the figure, the red dashed line stands for the case with $\sv=10^{-29}~\cms$ and $m_{\rm DM}=100$~MeV for dark matter annihilation to $\gamma\gamma$ channel. As we increase the cross-section to $\sv=10^{-28}~\cms$, the Ly-$\alpha$ coupling increases. For the $e^+e^-$ channel, we find a higher Ly-$\alpha$ coupling (yellow dot-dashed line) than that for the $\gamma\gamma$ channel (green dotted line) and consequently a higher absorptional amplitude during the dark ages. The same can also be seen as we go towards higher redshift values in Figure \ref{fig:21cm:a}.  

\begin{figure*}
    \begin{center}
        \subfloat[] {\includegraphics[width=3.2in,height=2in]{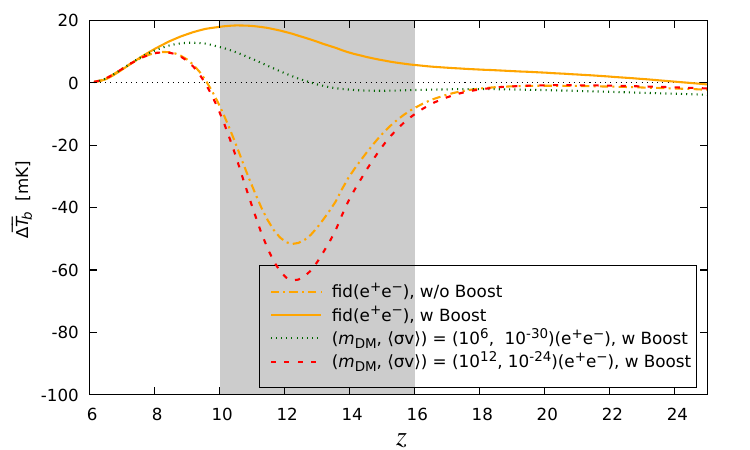}\label{fig:21cmB:a}}
        \subfloat[] {\qquad\includegraphics[width=3.2in,height=2in]{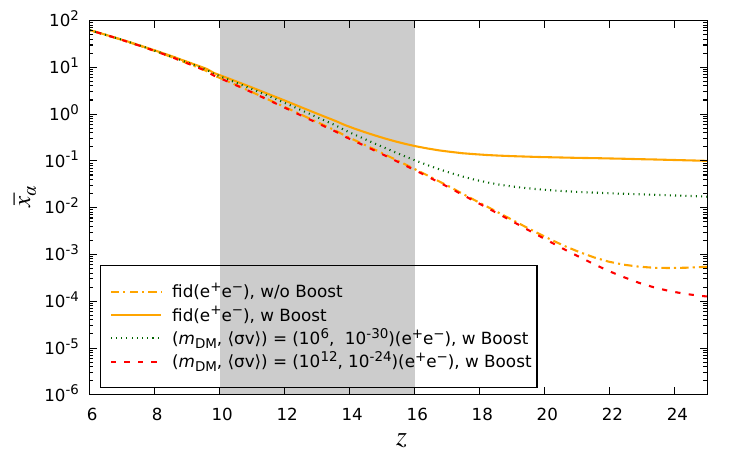}\label{fig:21cmB:b}}  
    \end{center}
    \caption{Left: Global 21-cm signal as a function of redshift. The dotted black horizontal line in the left figure is for reference--- represents the scenario with no 21-cm signal, i.e. no absorption and no emission ($\Delta T_b=0$). Right:  Global Ly-$\alpha$ coupling coefficient as a function of redshift. In both figures, `w/o boost' and `w boost' represent the cases without boost factor and with boost factor, respectively. The bracket, ($m_{\rm DM}$, $\sv$), represents the dark matter mass in `eV' and annihilation cross-section in `$\cms$' for corresponding cases. In these Figures, we compare how the structure formation boost affects the global 21-cm signal and Ly-$\alpha$ coupling. We have taken the $e^+e^-$ annihilation channel, $\DM  \DM \rightarrow e^+ e^-$, as an example to compare different scenarios. \label{fig:21cmB}}
\end{figure*}

In Figure \ref{fig:21cmB}, we also illustrate how the structure formation which enhances the dark matter annihilation can affect the Ly-$\alpha$ coupling and the global 21-cm signal, compared with the uniform dark matter density background without halos. In this figure, all the lines represent the case with $e^+e^-$ annihilation channel. The {yellow dot-dashed} line represents the fiducial case without boost (no structure formation and only the dark matter annihilation from the uniform dark matter background), while all the other solid line includes the structure formation boost. Here, it can be seen that the inclusion of the boost factor can have a significant effect on IGM evolution and, therefore, on the 21-cm signal. In Figure \ref{fig:21cmB:a} for the {yellow-coloured solid line}, we obtain an emission spectrum instead of absorptional spectra after the inclusion of the boost factor for all the considered redshift ranges. Therefore the energy deposition rate into the IGM increases as seen in equations (\ref{fcbz}) and  (\ref{eq:dep}). 
As a result, the gas temperature becomes larger than the CMB temperature giving an emission spectra. In this case, the spin temperature couples to the gas temperature early, which can be seen around a redshift of 24 in Figure \ref{fig:21cmB:a}. It is caused by a higher number of Ly-$\alpha$ photons from dark matter annihilation, i.e. a higher Ly-$\alpha$ coupling. This is shown by the {yellow solid line} in Figure \ref{fig:21cmB:b}. 
We also show two other parameter sets in this figure ({green-dotted} and {red-dashed} lines) for which we keep the constant ratio of $\langle \sigma\,v\rangle/m_{\rm DM}$ to study the degeneracy between different dark matter models and energy deposition efficiency, to be discussed further in section \ref{ssec:constraints_on_mass}.

\subsection{Thermal temperature noise}
\label{ssec:noise_generation}
We consider the system temperature noise associated with the instrumentation of the radio interferometers. Here, we assume the upcoming Square Kilometer Array (SKA) Low-frequency mode survey. For simplicity, we assume the noise is fully Gaussian, with zero mean and standard deviation in Fourier space \citep{2016MNRAS.459..863P},
\begin{align}
    & \sigma^2_N = \Delta^2_N d^2u, \label{eq:sigma_noise} \\
    & \Delta^2_N = \frac{T^2_{\rm sys}F^2}{B t_0 n_b} \label{eq:pk_noise},
\end{align}
where $du=2\pi \chi/L_{\rm box}$, B is the bandwidth of the frequency channel, $t_0$ is the integration time, $F$ is the field of view, $F=\lambda^2/A_d$, where $\lambda$ is the observed wavelength and $A_d=\pi(D_d/2)^2$ is the area of the single antenna dish, $D_d$ is the diameter of the single antenna dish. $T_{\rm sys}$ is the frequency-dependent system temperature, and $n_b$ is the number density of the baselines,
\begin{equation}
    n_b=\frac{N_d(N_d-1)}{2\pi (u^2_{\rm max}-u^2_{\rm min})},
\end{equation}
where $N_d$ is the number of antennas, $u_{\rm max}$ and $u_{\rm min}$ are defined using the maximum baseline $D_{\rm max}$ and the diameter of the single dish antenna, $u_{\rm max}=D_{\rm max}/\lambda$, and $u_{\rm min}=D_d/\lambda$.
In equation (\ref{eq:pk_noise}), we define the dimensionless power spectrum of the noise, which is used in the likelihood analysis with the power spectrum in section \ref{ssec:model_discrimination}. We add the random Gaussian noise with a variance of eq. (\ref{eq:sigma_noise}) to the brightness temperature images, $\Delta T_b=\mathcal{N}(0,\sigma_N)$.

In this paper, for specific computation, we assume SKA-Low to observe 21-cm line maps at $z=10, 13$ and 16 and adopt the experimental setup values from the SKA specifications \footnote{\href{https://www.skao.int/en/explore/telescopes/ska-low}{https://www.skao.int/en/explore/telescopes/ska-low}}. Therefore, we assume the numbers $N_d=512$, $D_{\rm max}= 74~ {\rm km}$. We assume that the integration time is $10^4$ hours in total, and the bandwidth $B$ is given by the frequency width corresponding to 1.5 Mpc/$h$ at each redshift. $B$ is roughly of order $100$ kHz and is comparable or wider than the current design of the SKA frequency channel widths.

\begin{figure*}
    \includegraphics[width=0.45\linewidth]{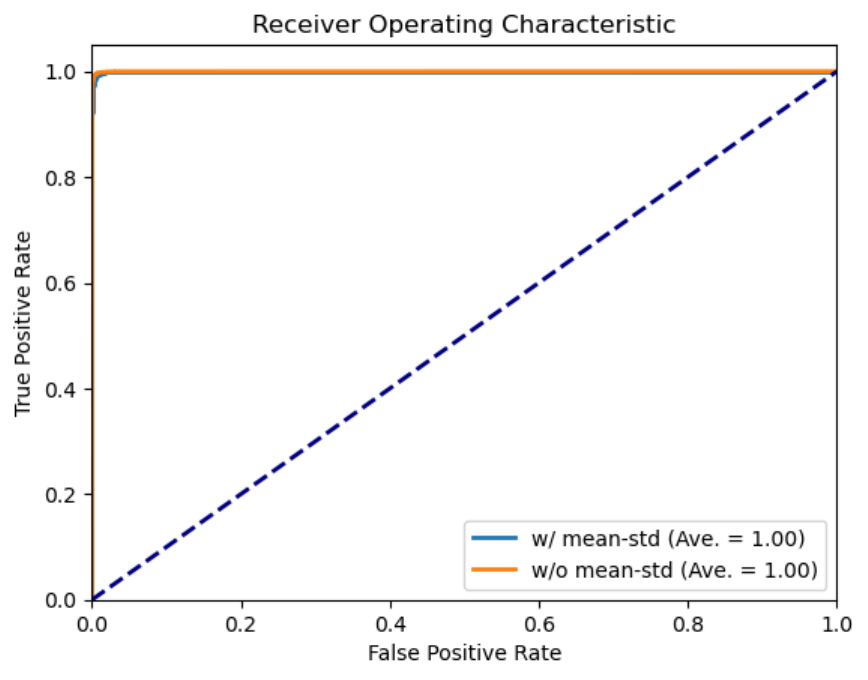}
    \includegraphics[width=0.45\linewidth]{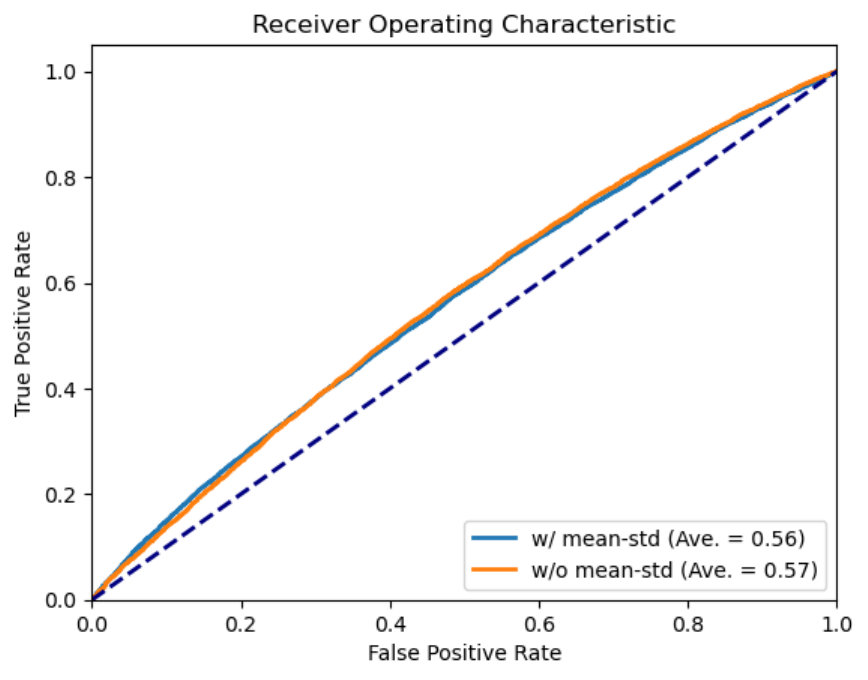}
    \caption{ROC for fiducial annihilation model at $z=10$ (left) and $\gamma \gamma$ channel with the cross section of $10^{-29}~\cms$ (right) \label{fig:roc}}
\end{figure*}

\subsection{model discrimination with CNN}
\label{ssec:model_discrimination}
We have used 25,600 images from the test sample, which are not used for training and validation processes. We first explore if the dark matter annihilation model can be discriminated from the non-annihilation scenario. Half of the test images are for the non-annihilation scenario, and the rest are from the annihilation scenario. The metric of the discrimination accuracy is computed using the AUC (Area Under Curve) of the ROC (Receiver Operating Characteristics) curve. The ROC is defined by the false positive rate and true positive rate. Here, we define the annihilation model as the \textit{positive}. Figure \ref{fig:roc} shows the ROC curves for two different annihilation models. The false positive rate (FPR=FP/(FP+TN): horizontal axis) is the number of non-annihilation scenario images mispredicted as the annihilation model divided by the total number of non-annihilation images, while the true positive rate (TPR=TP/(TP+FN)): vertical axis) is the number of annihilation images which is correctly predicted as the annihilation model divided by the total number of images being predicted as positive (annihilation). We consider the prediction to become positive when the prediction exceeds the given threshold value. The choice of the threshold is arbitrary but optimized so that the distance to the ROC curve from the diagonal line in the ROC plot is maximized or to maximize the F1 score defined as the harmonic average of the TPR and PRE,
\begin{equation}
F_1 = \frac{2}{{\rm TPR}^{-1}+{\rm PRE}^{-1}},
\end{equation}
where PRE stands for the precision, TP/(TP+FP), which means the fraction of the positive images correctly identified out of all the positive images. This $F_1$ score depends on the choice of the threshold. We take the threshold so that it maximizes the $F_1$ score.

Figure \ref{fig:roc} shows the ROC for two cases. The left panel represents the case where annihilation and non-annihilation models are well discriminated. Here we assume the $\ee$ annihilation channel and dark matter mass is 100MeV, annihilation cross-section $\sv=10^{-28} \cms$. The AUC value is 1.0, which means 100\% of the test images are correctly classified.
The right panel shows the case where annihilation and non-annihilation models are not well discriminated. We assume $\gamgam$ annihilation channel with $m_{\DM}=100$MeV, $\sv=10^{-29}\cms$. Here, we observe AUC=0.56, indicating that the classification is almost equivalent to the random drawing, and the CNN does not work properly at all.

The upper half of table \ref{tab:auc} shows the AUC values for dark matter models of mass $m_\DM=100$~MeV across various cross-sections. For the most conservative case, where we do not include any boost factor, we can expect to discriminate the non-annihilation scenario from the annihilation scenarios for $\sv>10^{-28}\cms$ at $z=10$ and 16 while the constraints are more stringent $\sv>10^{-29}\cms$ at $z=13$. Furthermore, if we include the substructure boost factor, the constraints become much tighter. For the $\ee$ channel, $\sv>10^{-30}\cms$ at $z=10$ and 16 and $\sv>10^{-31}\cms$ at $z=13$. On the other side, the constraint for the $\gamgam$ case sees only slight improvements. 
This is due to a very large mean free path of the injected photons. It exceeds the box size, and they can not be absorbed instantly into the plasma. The absorption of photons is efficient when their energy remains $\lesssim 10^{-1}$~keV for the considered box size \cite{Sun:2023}. The other way for these high energy photons, $\gtrsim$keV range, to get absorbed into the plasma is by getting redshifted in the $\lesssim 10^{-1}$~keV range due to the expansion of the Universe. The effect of structure formation on the dark matter annihilation becomes effective after a redshift of about 50. For example, if a 1~MeV photon is emitted by dark matter annihilation at a redshift of 50, it can maximally redshift to an energy of $10^2$~keV in the redshift range of interest--- i.e. it will freely move in the plasma. Therefore, even if the number density of emitted photons is enhanced due to the structure formation boost, their absorption efficiency into the plasma still remains nearly the same due to their large mean free path. Hence, we can not expect a significant improvement in the constraints for the photon case after the inclusion of the boost factor.  On the other hand, as discussed in section \ref{ssec:map_generation}, the energy deposition through the $\ee$ channel is most efficient around 100~MeV, and the energy deposition can take place nearly on the spot. Therefore, the effect of boost is significantly imprinted in the 21-cm signal and thus helps to isolate the annihilation signals for the $\ee$ channel.

\begin{table*}
    \begin{tabularx}{\textwidth}{c|XXX|XXX|XXX|XXX} 
        \hline \hline
        channel & \multicolumn{3}{c}{$e^+e^-$ (w/o Boost)} & \multicolumn{3}{c}{$\gamma\gamma$ (w/o Boost)} & \multicolumn{3}{c}{$e^+e^-$ (w/ Boost)}& \multicolumn{3}{c}{$\gamma\gamma$ (w/ Boost)} \\
        \hline
        redshift & 10 & 13 & 16 & 10 & 13 & 16 & 10 & 13 & 16 & 10 & 13 & 16 \\
        $\sv=10^{-28}$ [cm$^{3}$/s] & \cog{1.00} & \cog{1.00} & \cog{1.00}
                                    & \cog{0.99} & \cog{1.00} & \cog{1.00} 
                                    & \cog{1.00} & \cog{1.00} & \cog{1.00} 
                                    & \cog{1.00} & \cog{1.00} & \cog{1.00} \\
        $\sv=10^{-29}$ [cm$^{3}$/s] & \cor{0.64} & \cog{1.00} & \cor{0.78} 
                                    & \cor{0.51} & \cog{0.97} & \cor{0.56} 
                                    & \cog{1.00} & \cog{1.00} & \cog{1.00} 
                                    & \cor{0.57} & \cog{1.00} & \cor{0.71} \\
        $\sv=10^{-30}$ [cm$^{3}$/s] & $\cor{0.43}$ & $\cor{0.40}$ & $\cor{0.48}$ 
                                    & $\cor{0.44}$ & $\cor{0.24}$ & $\cor{0.48}$ 
                                    & \cog{1.00} & \cog{1.00} & \cog{1.00} 
                                    & \cor{0.43} & \cor{0.47} & \cor{0.29} \\
        $\sv=10^{-31}$ [cm$^{3}$/s] & $-$ & $-$ & $-$ 
                                    & $-$ & $-$ & $-$ 
                                    & \cor{0.79} & \cog{1.00} & \cor{0.83} 
                                    & $-$ & $-$ & $-$ \\
        $\sv=10^{-32}$ [cm$^{3}$/s] & $-$ & $-$ & $-$ 
                                    & $-$ & $-$ & $-$ 
                                    & \cor{0.44} & \cor{0.49} & \cor{0.34} 
                                    & $-$ & $-$ & $-$ \\
        \hline
    \end{tabularx}
    \begin{tabularx}{\textwidth}{c|XXX|XXX|XXX|XXX} 
        \hline
        channel & 
        \multicolumn{3}{c}{$e^+e^-$ (w/o Boost)} & 
        \multicolumn{3}{c}{$\gamma\gamma$ (w/o Boost)} & 
        \multicolumn{3}{c}{$e^+e^-$ (w/ Boost)} & 
        \multicolumn{3}{c}{$\gamma\gamma$ (w/ Boost)} \\
        \hline
        redshift & 10 & 13 & 16 & 
                   10 & 13 & 16 & 
                   10 & 13 & 16 & 
                   10 & 13 & 16 \\
        $\sv=10^{-28}$ [cm$^3$/s] & \cog{0.0094} & \cog{0.00} & \cor{0.85} & 
                                    \cor{0.65}   & \cog{0.00} & \cor{1.00} & 
                                    \cog{0.00}   & \cog{0.00} & \cor{0.72} & 
                                    \cor{0.061}  & \cog{0.00} & \cor{0.95} \\
        $\sv=10^{-29}$ [cm$^3$/s] & \cor{1.00}   & \cog{0.00} & \cor{1.00} & 
                                    \cor{1.00}   & \cog{0.0013} & \cor{1.00} & 
                                    \cog{0.00}   & \cog{0.00} & \cor{0.12} & 
                                    \cor{1.00}   & \cog{0.00} & \cor{1.00} \\
        $\sv=10^{-30}$ [cm$^3$/s] & $\cor{1.00}$ & $\cor{1.00}$ & $\cor{1.00}$ & 
                                    $\cor{1.00}$ & $\cor{1.00}$ & $\cor{1.00}$ & 
                                    \cog{0.001} & \cog{0.00} & \cor{0.82} & 
                                    \cor{1.00}  & \cor{1.00} & \cor{1.00} \\
        $\sv=10^{-31}$ [cm$^3$/s] & $-$ & $-$ & $-$ & 
                                    $-$ & $-$ & $-$ & 
                                    \cor{1.00} & \cog{0.00} & \cor{1.00} & 
                                    $-$ & $-$ & $-$ \\
        $\sv=10^{-32}$ [cm$^3$/s] & $-$ & $-$ & $-$ & 
                                    $-$ & $-$ & $-$ & 
                                    \cor{1.00} & \cor{1.00} & \cor{1.00} & 
                                    $-$ & $-$ & $-$ \\
        \hline \hline
    \end{tabularx}
    \caption{Constraints on the dark matter cross-section parameter for the dark matter mass $m_{\DM}=100$ MeV. The upper table shows the AUC values for discriminating the dark matter annihilation scenarios from the no annihilation using CNN. The lower table represents the p-values for discriminating the annihilation and non-annihilation scenarios based on the power spectrum analysis.
    \label{tab:auc}
    }
\end{table*}
%
%
%
%

\subsection{Comparison to traditional $P(k)$ analysis}
\label{ssec:model_discrimination}
In this section, we compare the dark matter model constraints by CNN with the one from standard power spectrum analysis and demonstrate that the image-based analysis outperforms the traditional power spectrum analysis. We compute the power spectra for each 3-dimensional image based on eq. (\ref{eq:powersp}), but adding the noise given by eq. (\ref{eq:sigma_noise}), 
\begin{equation}
    \tilde{\Delta}_{21}^2 = \Delta_{21}^2 + \Delta^2_N.
\end{equation}
\begin{figure}
    \includegraphics[width=\linewidth]{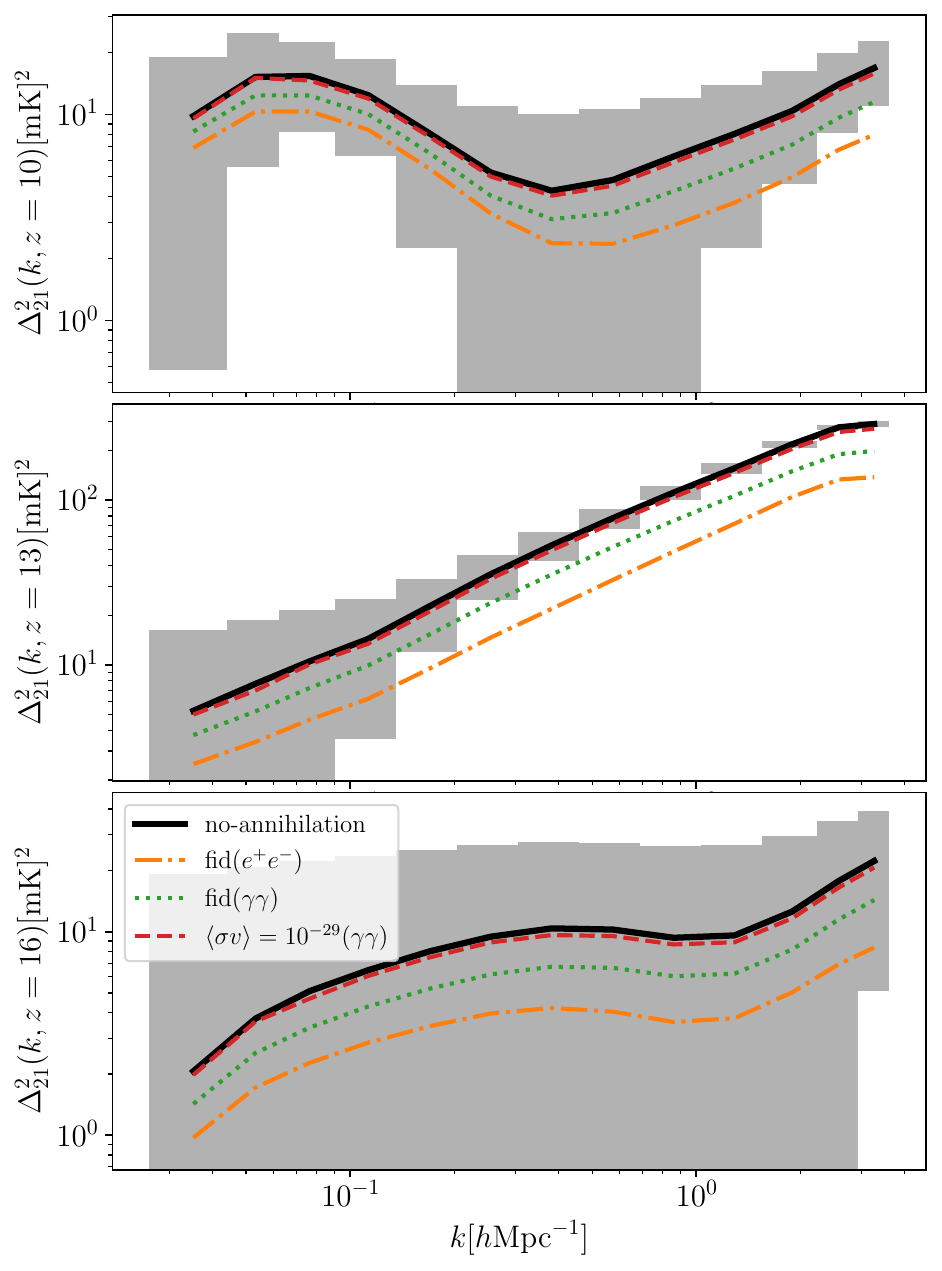}
    \caption{Power spectra of the 21-cm line at $z=10, 13$ and 16 from top to bottom. The shaded region is the expected error for cosmic variance equivalent to 300 Mpc/$h$ volume and SKA-Low noise level of $10^4$ hours integration. 
        \label{fig:pk21}}
\end{figure}
The maximum and minimum $k$ should be determined by the pixel size of the image and simulation box size, respectively. Namely, they are exactly the same as the image data used for the CNN. Therefore, there are no excesses and deficiencies induced by the difference in the configuration of the data, which ensures our comparison is fair. We compute the $\chi^2$ for power spectrum averaged over 1,000 realizations as 
\begin{align}
    &\chi^2
    =
    \sum_{\bm{k}}
    [\delta \Delta^2_{21}(\bm{k}) ]
    {\bf C}^{-1}
    [\delta \Delta^2_{21}(\bm{k})]^t, \\
    &
    \delta \Delta^2_{21}(\bm{k})
     = \Delta_{21, {\rm ann.}}^2(\bm{k})-\Delta_{21, {\rm non-ann.}}^2(\bm{k}),
\end{align}
where ${\bf C}$ is the covariance matrix computed from the 1,000 realizations of the images for the conventional $\Lambda$CDM model without dark matter annihilation, $\Delta_{21,{\rm ann.}}$ and $\Delta_{21, {\rm non-ann.}}$ are the brightness temperature power spectra for annihilation and non-annihilation model respectively.
In Figure \ref{fig:pk21}, we show the 21-cm signal power spectrum. Here, we use the power spectrum definition given by \eqref{eq:powersp}.
Since this is the dimensionless power spectrum times the global average temperature squared, the overall amplitude is determined by both the amplitude of the global signal and the fluctuation amplitude. 
There exists a non-trivial k-dependence in our parameters, such as in Ly-$\alpha$ coupling coefficient $x_{\alpha}$ and in neutral hydrogen fraction $x_{\rm HI}$, which results in the non-trivial k-dependence in the power spectra. The error bar is due to the thermal noise defined in section \ref{ssec:noise_generation} and the cosmic variance due to the finite simulation box size. This roughly corresponds to the sky area of 3 square degrees, which is readily covered by the SKA. 

The bottom half of table \ref{tab:auc} presents the constraints from the power spectrum analysis. We compute the p-value from the $\chi^2$ distribution with the degree of freedom being 13.
In the case of no boost factor, we cannot discriminate the annihilation scenarios from the non-annihilation scenario if $\sv \lesssim 10^{-31}$, regardless of the annihilation channel.

\subsection{TeV and MeV models}
\label{ssec:constraints_on_mass}

\begin{table}
    \begin{tabularx}{\linewidth}{c|c|XXX} 
        \hline \hline
        \, & redshift & 10 & 13 & 16 \\
        \hline
        $m_\DM=$1MeV & $\sv=10^{-24}$ [cm$^{3}$/s] & \cog{0.93} &\cog{1.00} & \cog{0.99} \\
        \,   & $\sv=10^{-25}$ [cm$^{3}$/s] & \cor{0.48} & \cor{0.87} & \cor{0.50}  \\
                                    \hline
        $m_\DM=$1TeV & $\sv=10^{-31}$ [cm$^{3}$/s] & \cog{1.00} & \cog{1.00} & \cog{1.00}\\
        \,   & $\sv=10^{-32}$ [cm$^{3}$/s] & \cog{0.92} & \cog{1.00} & \cor{0.84}\\
        \,   & $\sv=10^{-33}$ [cm$^{3}$/s] & \cor{0.45} & \cor{0.43} & \cor{0.45}\\
        \hline \hline
    \end{tabularx}
    \caption{
    AUC values for discriminating the annihilation scenarios, through $\ee$ channel with boost factor, from the no-annihilation scenarios. 
    \label{tab:auc_MeVTeV}
    }
\end{table}

So far, we have discussed the 21 cm signals for the dark matter mass of $m_\DM=100$ MeV for which the energy deposition from $\DM \DM \rightarrow \ee$ into the gas is most efficient for the redshift range of our interest. In this section, we illustrate our findings for other dark matter mass ranges, more concretely for MeV and TeV masses. For $m_\DM=1$ MeV and 1 TeV with the annihilation channel $\DM \DM \rightarrow \ee$, the current tightest upper bound on $\sv$ notably from the CMB is, respectively, of order a few times $10^{-30}\cms$ and of order $10^{-24}\cms$ \cite{2016PhRvD..93b3527S}. We demonstrate here that the CNN method can probe a comparable or even smaller annihilation cross-section, depending on the probed redshift range, for which 21 signals can distinguish the annihilation scenarios from the non-annihilation scenarios.

When we consider constraining the cross-section and mass of the dark matter model simultaneously, the most serious concern is that there is a degeneracy between them if we keep $f_c$ to a constant, $\approx1/3$, corresponding to the instantaneous deposition (or on-the-spot) case. In such a case, we cannot distinguish between different dark matter models where  $\langle \sigma\,v\rangle/m_\DM$ ratio remains the same. Therefore, introducing the non-constant energy deposition efficiency factor, $f_c(z,m_{DM})$, beyond a simple on-the-spot approximation can help resolve the degeneracy and study the different dark matter models in the observational data. 

For the case of $m_\DM=1$~MeV ({green dotted line}) in Figure \ref{fig:21cmB:a}, the dark matter density increases compared with $m_\DM=10^2$~MeV denoted by the {yellow solid line}, as the total energy density of dark matter should remain the same. This results in the enhancement of the annihilation rate, which is proportional to the dark matter density squared. However, this case results in less energetic electrons and the energy deposition efficiency factor becomes $\mathcal{O}(10^1)$ smaller compared with $100$ MeV case. The two main processes for the energy deposition into the gas are collisional ionizations and collisional excitations for both the electrons and positrons. Even though the inverse Compton scattering still stays the dominating energy loss process for electrons, the upscattered CMB photons energy remains less than $\sim0.1$~eV in the redshift range of interest. Therefore, overall, collisional ionizations and excitations are the primary modes for energy deposition into the IGM \cite{Valds:2010, Evoli:2012, Liu:2016}. This results in less heating of the gas--- i.e. smaller emission spectra for the global 21-cm signal. The corresponding Ly-$\alpha$ coupling is shown in Figure \ref{fig:21cmB:b}. 

For the case of $m_\DM=1$~TeV ({red dashed line}) with $\langle \sigma\,v\rangle=10^{-24} \cms$ in Figure \ref{fig:21cmB:a}, as we increase the dark matter mass, the number density of dark matter particles decreases resulting in a smaller annihilation rate. Moreover, the larger dark matter mass results in more energetic electrons/positrons. As the emitted electrons for this case can upscatter CMB photons above $\sim10^{10}$~eV in the redshift of interest, the IGM becomes mostly transparent and the photons are not so efficiently absorbed, leading to a smaller heating of IGM compared to the case of $m_\DM=10^2$~MeV case \cite{Valds:2010, Evoli:2012, Liu:2016}. Consequently, the Ly-$\alpha$ coupling also decreases, as shown by the {red dashed} line in Figure \ref{fig:21cmB:b}.

We apply the general form of the energy deposition rate $f_c(m_{DM},z)$ and generate the images for 1MeV and 1TeV with various cross-section rates. Table \ref{tab:auc_MeVTeV} summarizes the AUC values constrained by the CNN model. For the 1~MeV case, we can constrain the cross-section fairly stringently of order $\sv>10^{-32} \cms$.
For $m_\DM=1$~TeV, the required minimum value is of order $\sv > 10^{-24} \cms$ 
or even smaller for some redshift range, for which the 21-cm observation can probe the effects of dark matter annihilation distinguishable from the non-annihilation scenarios. Hence our CNN analysis on the 21-cm signals can be more sensitive to the dark matter annihilation signatures than the CMB and X-ray observations whose current upper bounds are of order $\sv= 10^{-30}\cms$ and $10^{-24} \cms$ for $m_\DM=1$ MeV and 1 TeV, respectively 
\cite{Cirelli:2023tnx,2016PhRvD..93b3527S}.

We also note that the systematic effects associated with the observation are different for the different observational probes. Therefore, although the constraints on $m_\DM=1$TeV are marginally comparable to those from the CMB, independent constraints from other observational probes are still valuable, and the currently expected constraints from the 21-cm line map can be readily scaled by the sky area we observe as $1/\sqrt{f_{\rm sky}}$.

\section{Summary}
\label{sec:summary}
In this paper, we investigate the influence of dark matter annihilation on the 21-cm signal during the epochs of cosmic dawn and reionization, focusing on the energy injection into the intergalactic medium (IGM) and its effects on the observable 21-cm brightness temperature. Using a combination of semi-numerical simulations with the \texttt{21cmFAST} code and machine learning techniques, particularly Convolutional Neural Networks (CNNs), we explore how dark matter annihilation effects can be detected by differentiating between annihilating and non-annihilating dark matter models. Our work emphasizes the potential of CNNs to enhance the detection and characterization of dark matter effects on the 21-cm signal, especially during the early stages of cosmic structure formation.

Our simulations focus on two primary annihilation channels: photon-photon ($\gamgam$) and electron-positron ($\ee$), exploring a wide range of dark matter masses and cross-sections. These channels represent different forms of energy injection into the IGM, which in turn influences the evolution of the 21-cm signal. By modifying the \texttt{21cmFAST} code to incorporate the effects of dark matter annihilation, we generate three-dimensional maps of the 21-cm brightness temperature on large scales, allowing us to simulate the cosmological scenarios both with and without dark matter annihilation.

The main challenge in analyzing the 21-cm signal lies in distinguishing the dark matter annihilation effects in the presence of the fluctuations expected from the conventional $\Lambda$CDM model without annihilation. Previous studies predominantly have focused on summary statistics, such as the power spectrum, which may fail to capture the complex spatial patterns present in the full three-dimensional data. To address this issue, we applied CNNs, which have shown considerable success in other areas of astrophysics, to analyze the spatial patterns in the 21-cm maps. Our results demonstrate that CNNs can outperform those traditional summary statistics approaches in distinguishing between the annihilation and no-annihilation models. Even in cases where the differences between these models are subtle, when the annihilation cross-sections are small, the CNN is able to effectively discriminate them.

We further explore the impact of structure formation on dark matter annihilation. As the Universe evolves, dark matter halos begin to form, enhancing the density in localized regions and increasing the dark matter annihilation rate. We calculate numerically the 21-cm signals in the presence of dark matter annihilation effects enhanced by such halo formation as well as subhalos inside host halos and demonstrate the potential power of machine learning to distinguish those dark matter annihilation models from the non-annihilation models.

Our analysis indicates that at higher redshifts, the IGM is relatively cold, and the 21-cm signal appears as a strong absorption feature. As dark matter annihilation proceeds and the IGM heats up, the amplitude of this absorption signal decreases and, in some cases, transitions into emission. This effect is particularly evident at lower redshifts ($z \simeq 10$), where the structure formation boost becomes significant. The presence of this emission feature, driven by the annihilation, could serve as a clear indicator of the underlying dark matter properties, including the mass and annihilation cross-section.

We also quantify how the inclusion of noise from observational systems, specifically from the upcoming Square Kilometer Array (SKA), could affect the detectability of dark matter annihilation signals. By modelling the noise expected in future 21-cm observations, we demonstrate that even in the presence of observational limitations, CNNs are capable of distinguishing between annihilation and non-annihilation models. We expect that SKA, with its ability to observe the 21-cm signal across a range of redshifts, will provide invaluable data for probing dark matter properties such as their mass and annihilation rate.

A key finding of our study is the importance of incorporating detailed modelling of energy deposition processes into analyses of dark matter effects on the 21-cm signal. Different dark matter properties, such as different decay channels, mass, and annihilation rate,  deposit energy into the IGM through various mechanisms, including heating, ionization, and Ly-$\alpha$ coupling, each of which leaves distinct imprints on the 21-cm signal. By including these processes in our simulations, we are able to provide more accurate predictions for the global and the fluctuations of 21-cm signals, allowing us to better constrain dark matter properties.

Our results show that 21cn signals can detect the effects of the dark matter annihilation for the dark matter masses $m_\DM=1$ MeV, 100 MeV and 1 TeV respectively, with the annihilation cross-sections $\sv \simeq 10^{-32}~\cms, 10^{-31}~\cms$ and $10^{-24}~\cms$. This indicates the potential of CNN analysis for the forthcoming 21-cm signals which can probe the dark matter parameter range which cannot be probed by other observations such as the CMB whose current upper bounds are of order $\sv= 10^{-30}~\cms,10^{-29}~\cms$ and $10^{-24}~\cms$ respectively for $m_\DM=1$ MeV, 100 MeV and 1 TeV \cite{Cirelli:2023tnx,2016PhRvD..93b3527S}.

Future work will focus on refining these methods and applying them to real observational data from the SKA and other upcoming 21-cm experiments. With improved sensitivity and resolution, these experiments will allow us to place even tighter constraints on dark matter properties and explore a wider range of dark matter models.

\begin{acknowledgments}
We thank Kazunori Kohri and Yin Li for the useful discussions. This work is in part supported by JSPS Kakenhi Grants JP21H05454, JP21K03625 and JP23H00108, JSPS International Leading Research, JP22K21349, and by Center of Quantum Cosmo Theoretical Physics (NSFC grant number 12347103).
\end{acknowledgments}



\bibliography{bibtex, ref}

\begin{thebibliography}{53}%
\makeatletter
\providecommand \@ifxundefined [1]{%
 \@ifx{#1\undefined}
}%
\providecommand \@ifnum [1]{%
 \ifnum #1\expandafter \@firstoftwo
 \else \expandafter \@secondoftwo
 \fi
}%
\providecommand \@ifx [1]{%
 \ifx #1\expandafter \@firstoftwo
 \else \expandafter \@secondoftwo
 \fi
}%
\providecommand \natexlab [1]{#1}%
\providecommand \enquote  [1]{``#1''}%
\providecommand \bibnamefont  [1]{#1}%
\providecommand \bibfnamefont [1]{#1}%
\providecommand \citenamefont [1]{#1}%
\providecommand \href@noop [0]{\@secondoftwo}%
\providecommand \href [0]{\begingroup \@sanitize@url \@href}%
\providecommand \@href[1]{\@@startlink{#1}\@@href}%
\providecommand \@@href[1]{\endgroup#1\@@endlink}%
\providecommand \@sanitize@url [0]{\catcode `\\12\catcode `\$12\catcode
  `\&12\catcode `\#12\catcode `\^12\catcode `\_12\catcode `\%12\relax}%
\providecommand \@@startlink[1]{}%
\providecommand \@@endlink[0]{}%
\providecommand \url  [0]{\begingroup\@sanitize@url \@url }%
\providecommand \@url [1]{\endgroup\@href {#1}{\urlprefix }}%
\providecommand \urlprefix  [0]{URL }%
\providecommand \Eprint [0]{\href }%
\providecommand \doibase [0]{https://doi.org/}%
\providecommand \selectlanguage [0]{\@gobble}%
\providecommand \bibinfo  [0]{\@secondoftwo}%
\providecommand \bibfield  [0]{\@secondoftwo}%
\providecommand \translation [1]{[#1]}%
\providecommand \BibitemOpen [0]{}%
\providecommand \bibitemStop [0]{}%
\providecommand \bibitemNoStop [0]{.\EOS\space}%
\providecommand \EOS [0]{\spacefactor3000\relax}%
\providecommand \BibitemShut  [1]{\csname bibitem#1\endcsname}%
\let\auto@bib@innerbib\@empty
\bibitem [{\citenamefont {{Planck Collaboration et al.}}(2020)}]{Planck:2018}%
  \BibitemOpen
  \bibfield  {author} {\bibinfo {author} {\bibnamefont {{Planck Collaboration
  et al.}}},\ }\bibfield  {title} {\bibinfo {title} {Planck 2018 results. vi.
  cosmological parameters},\ }\href
  {https://doi.org/10.1051/0004-6361/201833910} {\bibfield  {journal} {\bibinfo
   {journal} {A\&A}\ }\textbf {\bibinfo {volume} {641}},\ \bibinfo {pages} {A6}
  (\bibinfo {year} {2020})}\BibitemShut {NoStop}%
\bibitem [{\citenamefont {Boddy}\ \emph {et~al.}(2022)\citenamefont {Boddy}
  \emph {et~al.}}]{Boddy:2022knd}%
  \BibitemOpen
  \bibfield  {author} {\bibinfo {author} {\bibfnamefont {K.~K.}\ \bibnamefont
  {Boddy}} \emph {et~al.},\ }\bibfield  {title} {\bibinfo {title}
  {{Snowmass2021 theory frontier white paper: Astrophysical and cosmological
  probes of dark matter}},\ }\href
  {https://doi.org/10.1016/j.jheap.2022.06.005} {\bibfield  {journal} {\bibinfo
   {journal} {JHEAp}\ }\textbf {\bibinfo {volume} {35}},\ \bibinfo {pages}
  {112} (\bibinfo {year} {2022})},\ \Eprint {https://arxiv.org/abs/2203.06380}
  {arXiv:2203.06380 [hep-ph]} \BibitemShut {NoStop}%
\bibitem [{\citenamefont {Cooley}\ \emph {et~al.}(2022)\citenamefont {Cooley}
  \emph {et~al.}}]{Cooley:2022ufh}%
  \BibitemOpen
  \bibfield  {author} {\bibinfo {author} {\bibfnamefont {J.}~\bibnamefont
  {Cooley}} \emph {et~al.},\ }\bibfield  {title} {\bibinfo {title} {{Report of
  the Topical Group on Particle Dark Matter for Snowmass 2021}},\ }\href@noop
  {} {\  (\bibinfo {year} {2022})},\ \Eprint {https://arxiv.org/abs/2209.07426}
  {arXiv:2209.07426 [hep-ph]} \BibitemShut {NoStop}%
\bibitem [{\citenamefont {Feng}(2010)}]{Feng:2010}%
  \BibitemOpen
  \bibfield  {author} {\bibinfo {author} {\bibfnamefont {J.~L.}\ \bibnamefont
  {Feng}},\ }\bibfield  {title} {\bibinfo {title} {Dark matter candidates from
  particle physics and methods of detection},\ }\href
  {https://doi.org/10.1146/annurev-astro-082708-101659} {\bibfield  {journal}
  {\bibinfo  {journal} {Annual Review of Astronomy and Astrophysics}\ }\textbf
  {\bibinfo {volume} {48}},\ \bibinfo {pages} {495–545} (\bibinfo {year}
  {2010})}\BibitemShut {NoStop}%
\bibitem [{\citenamefont {Liu}\ \emph {et~al.}(2016)\citenamefont {Liu},
  \citenamefont {Slatyer},\ and\ \citenamefont {Zavala}}]{Liu:2016}%
  \BibitemOpen
  \bibfield  {author} {\bibinfo {author} {\bibfnamefont {H.}~\bibnamefont
  {Liu}}, \bibinfo {author} {\bibfnamefont {T.~R.}\ \bibnamefont {Slatyer}},\
  and\ \bibinfo {author} {\bibfnamefont {J.}~\bibnamefont {Zavala}},\
  }\bibfield  {title} {\bibinfo {title} {Contributions to cosmic reionization
  from dark matter annihilation and decay},\ }\bibfield  {journal} {\bibinfo
  {journal} {Physical Review D}\ }\textbf {\bibinfo {volume} {94}},\ \href
  {https://doi.org/10.1103/physrevd.94.063507} {10.1103/physrevd.94.063507}
  (\bibinfo {year} {2016})\BibitemShut {NoStop}%
\bibitem [{\citenamefont {Liu}\ \emph {et~al.}(2020)\citenamefont {Liu},
  \citenamefont {Ridgway},\ and\ \citenamefont {Slatyer}}]{Liu_2020}%
  \BibitemOpen
  \bibfield  {author} {\bibinfo {author} {\bibfnamefont {H.}~\bibnamefont
  {Liu}}, \bibinfo {author} {\bibfnamefont {G.~W.}\ \bibnamefont {Ridgway}},\
  and\ \bibinfo {author} {\bibfnamefont {T.~R.}\ \bibnamefont {Slatyer}},\
  }\bibfield  {title} {\bibinfo {title} {Code package for calculating modified
  cosmic ionization and thermal histories with dark matter and other exotic
  energy injections},\ }\bibfield  {journal} {\bibinfo  {journal} {Phys. Rev.
  D}\ }\textbf {\bibinfo {volume} {101}},\ \href
  {https://doi.org/10.1103/physrevd.101.023530} {10.1103/physrevd.101.023530}
  (\bibinfo {year} {2020})\BibitemShut {NoStop}%
\bibitem [{\citenamefont {D'Amico}\ \emph {et~al.}(2018)\citenamefont
  {D'Amico}, \citenamefont {Panci},\ and\ \citenamefont
  {Strumia}}]{Amico:2018}%
  \BibitemOpen
  \bibfield  {author} {\bibinfo {author} {\bibfnamefont {G.}~\bibnamefont
  {D'Amico}}, \bibinfo {author} {\bibfnamefont {P.}~\bibnamefont {Panci}},\
  and\ \bibinfo {author} {\bibfnamefont {A.}~\bibnamefont {Strumia}},\
  }\bibfield  {title} {\bibinfo {title} {Bounds on dark-matter annihilations
  from 21-cm data},\ }\href {https://doi.org/10.1103/PhysRevLett.121.011103}
  {\bibfield  {journal} {\bibinfo  {journal} {Phys. Rev. Lett.}\ }\textbf
  {\bibinfo {volume} {121}},\ \bibinfo {pages} {011103} (\bibinfo {year}
  {2018})}\BibitemShut {NoStop}%
\bibitem [{\citenamefont {Mesinger}\ \emph {et~al.}(2011)\citenamefont
  {Mesinger}, \citenamefont {Furlanetto},\ and\ \citenamefont
  {Cen}}]{Mesinger:2011FS}%
  \BibitemOpen
  \bibfield  {author} {\bibinfo {author} {\bibfnamefont {A.}~\bibnamefont
  {Mesinger}}, \bibinfo {author} {\bibfnamefont {S.}~\bibnamefont
  {Furlanetto}},\ and\ \bibinfo {author} {\bibfnamefont {R.}~\bibnamefont
  {Cen}},\ }\bibfield  {title} {\bibinfo {title} {{21cmfast: a fast,
  seminumerical simulation of the high-redshift 21-cm signal}},\ }\href
  {https://doi.org/10.1111/j.1365-2966.2010.17731.x} {\bibfield  {journal}
  {\bibinfo  {journal} {MNRAS}\ }\textbf {\bibinfo {volume} {411}},\ \bibinfo
  {pages} {955} (\bibinfo {year} {2011})}\BibitemShut {NoStop}%
\bibitem [{\citenamefont {Murakami}\ and\ \citenamefont
  {Nishizawa}(2020)}]{Murakami:2020}%
  \BibitemOpen
  \bibfield  {author} {\bibinfo {author} {\bibfnamefont {K.}~\bibnamefont
  {Murakami}}\ and\ \bibinfo {author} {\bibfnamefont {A.~J.}\ \bibnamefont
  {Nishizawa}},\ }\href {https://arxiv.org/abs/2012.03778} {\bibinfo {title}
  {Identifying cosmological information in a deep neural network}} (\bibinfo
  {year} {2020}),\ \Eprint {https://arxiv.org/abs/2012.03778} {arXiv:2012.03778
  [astro-ph.CO]} \BibitemShut {NoStop}%
\bibitem [{\citenamefont {Sabiu}\ \emph {et~al.}(2022)\citenamefont {Sabiu},
  \citenamefont {Kadota}, \citenamefont {Asorey},\ and\ \citenamefont
  {Park}}]{Sabiu:2022}%
  \BibitemOpen
  \bibfield  {author} {\bibinfo {author} {\bibfnamefont {C.~G.}\ \bibnamefont
  {Sabiu}}, \bibinfo {author} {\bibfnamefont {K.}~\bibnamefont {Kadota}},
  \bibinfo {author} {\bibfnamefont {J.}~\bibnamefont {Asorey}},\ and\ \bibinfo
  {author} {\bibfnamefont {I.}~\bibnamefont {Park}},\ }\bibfield  {title}
  {\bibinfo {title} {Probing ultra-light axion dark matter from 21 cm
  tomography using convolutional neural networks},\ }\href
  {https://doi.org/10.1088/1475-7516/2022/01/020} {\bibfield  {journal}
  {\bibinfo  {journal} {Journal of Cosmology and Astroparticle Physics}\
  }\textbf {\bibinfo {volume} {2022}}\bibinfo  {number} { (01)},\ \bibinfo
  {pages} {020}}\BibitemShut {NoStop}%
\bibitem [{\citenamefont {Murakami}\ \emph
  {et~al.}(2024{\natexlab{a}})\citenamefont {Murakami}, \citenamefont {Ocampo},
  \citenamefont {Nesseris}, \citenamefont {Nishizawa},\ and\ \citenamefont
  {Kuroyanagi}}]{Murakami:2024}%
  \BibitemOpen
\bibfield  {number} {  }\bibfield  {author} {\bibinfo {author} {\bibfnamefont
  {K.}~\bibnamefont {Murakami}}, \bibinfo {author} {\bibfnamefont
  {I.}~\bibnamefont {Ocampo}}, \bibinfo {author} {\bibfnamefont
  {S.}~\bibnamefont {Nesseris}}, \bibinfo {author} {\bibfnamefont {A.~J.}\
  \bibnamefont {Nishizawa}},\ and\ \bibinfo {author} {\bibfnamefont
  {S.}~\bibnamefont {Kuroyanagi}},\ }\bibfield  {title} {\bibinfo {title}
  {Nonlinearity-free prediction of the growth-rate $f\sigma_8$ using
  convolutional neural networks},\ }\bibfield  {journal} {\bibinfo  {journal}
  {Physical Review D}\ }\textbf {\bibinfo {volume} {110}},\ \href
  {https://doi.org/10.1103/physrevd.110.023525} {10.1103/physrevd.110.023525}
  (\bibinfo {year} {2024}{\natexlab{a}})\BibitemShut {NoStop}%
\bibitem [{\citenamefont {Murakami}\ \emph
  {et~al.}(2024{\natexlab{b}})\citenamefont {Murakami}, \citenamefont {Kadota},
  \citenamefont {Nishizawa}, \citenamefont {Nagamine},\ and\ \citenamefont
  {Shimizu}}]{Murakami::2024}%
  \BibitemOpen
  \bibfield  {author} {\bibinfo {author} {\bibfnamefont {K.}~\bibnamefont
  {Murakami}}, \bibinfo {author} {\bibfnamefont {K.}~\bibnamefont {Kadota}},
  \bibinfo {author} {\bibfnamefont {A.~J.}\ \bibnamefont {Nishizawa}}, \bibinfo
  {author} {\bibfnamefont {K.}~\bibnamefont {Nagamine}},\ and\ \bibinfo
  {author} {\bibfnamefont {I.}~\bibnamefont {Shimizu}},\ }\bibfield  {title}
  {\bibinfo {title} {Differentiating warm dark matter models through 21-cm line
  intensity mapping: A convolutional neural network approach},\ }\bibfield
  {journal} {\bibinfo  {journal} {Physical Review D}\ }\textbf {\bibinfo
  {volume} {110}},\ \href {https://doi.org/10.1103/physrevd.110.023526}
  {10.1103/physrevd.110.023526} (\bibinfo {year}
  {2024}{\natexlab{b}})\BibitemShut {NoStop}%
\bibitem [{\citenamefont {Gillet}\ \emph {et~al.}(2019)\citenamefont {Gillet},
  \citenamefont {Mesinger}, \citenamefont {Greig}, \citenamefont {Liu},\ and\
  \citenamefont {Ucci}}]{Gillet:2019}%
  \BibitemOpen
  \bibfield  {author} {\bibinfo {author} {\bibfnamefont {N.}~\bibnamefont
  {Gillet}}, \bibinfo {author} {\bibfnamefont {A.}~\bibnamefont {Mesinger}},
  \bibinfo {author} {\bibfnamefont {B.}~\bibnamefont {Greig}}, \bibinfo
  {author} {\bibfnamefont {A.}~\bibnamefont {Liu}},\ and\ \bibinfo {author}
  {\bibfnamefont {G.}~\bibnamefont {Ucci}},\ }\bibfield  {title} {\bibinfo
  {title} {Deep learning from 21-cm tomography of the cosmic dawn and
  reionization},\ }\bibfield  {journal} {\bibinfo  {journal} {Monthly Notices
  of the Royal Astronomical Society}\ }\href
  {https://doi.org/10.1093/mnras/stz010} {10.1093/mnras/stz010} (\bibinfo
  {year} {2019})\BibitemShut {NoStop}%
\bibitem [{\citenamefont {Villanueva-Domingo}\ and\ \citenamefont
  {Villaescusa-Navarro}(2021)}]{Villanueva:2021}%
  \BibitemOpen
  \bibfield  {author} {\bibinfo {author} {\bibfnamefont {P.}~\bibnamefont
  {Villanueva-Domingo}}\ and\ \bibinfo {author} {\bibfnamefont
  {F.}~\bibnamefont {Villaescusa-Navarro}},\ }\bibfield  {title} {\bibinfo
  {title} {Removing astrophysics in 21 cm maps with neural networks},\ }\href
  {https://doi.org/10.3847/1538-4357/abd245} {\bibfield  {journal} {\bibinfo
  {journal} {The Astrophysical Journal}\ }\textbf {\bibinfo {volume} {907}},\
  \bibinfo {pages} {44} (\bibinfo {year} {2021})}\BibitemShut {NoStop}%
\bibitem [{\citenamefont {Zhu}\ \emph {et~al.}(2019)\citenamefont {Zhu},
  \citenamefont {Dai}, \citenamefont {Bian}, \citenamefont {Chen},
  \citenamefont {Chen},\ and\ \citenamefont {Hu}}]{Zhu:2019}%
  \BibitemOpen
  \bibfield  {author} {\bibinfo {author} {\bibfnamefont {X.-P.}\ \bibnamefont
  {Zhu}}, \bibinfo {author} {\bibfnamefont {J.-M.}\ \bibnamefont {Dai}},
  \bibinfo {author} {\bibfnamefont {C.-J.}\ \bibnamefont {Bian}}, \bibinfo
  {author} {\bibfnamefont {Y.}~\bibnamefont {Chen}}, \bibinfo {author}
  {\bibfnamefont {S.}~\bibnamefont {Chen}},\ and\ \bibinfo {author}
  {\bibfnamefont {C.}~\bibnamefont {Hu}},\ }\bibfield  {title} {\bibinfo
  {title} {Galaxy morphology classification with deep convolutional neural
  networks},\ }\bibfield  {journal} {\bibinfo  {journal} {Astrophysics and
  Space Science}\ }\textbf {\bibinfo {volume} {364}},\ \href
  {https://doi.org/10.1007/s10509-019-3540-1} {10.1007/s10509-019-3540-1}
  (\bibinfo {year} {2019})\BibitemShut {NoStop}%
\bibitem [{\citenamefont {Cheng}\ \emph {et~al.}(2021)\citenamefont {Cheng}
  \emph {et~al.}}]{Cheng:2021}%
  \BibitemOpen
  \bibfield  {author} {\bibinfo {author} {\bibfnamefont {T.-Y.}\ \bibnamefont
  {Cheng}} \emph {et~al.},\ }\bibfield  {title} {\bibinfo {title} {{Galaxy
  morphological classification catalogue of the Dark Energy Survey Year 3 data
  with convolutional neural networks}},\ }\href
  {https://doi.org/10.1093/mnras/stab2142} {\bibfield  {journal} {\bibinfo
  {journal} {Monthly Notices of the Royal Astronomical Society}\ }\textbf
  {\bibinfo {volume} {507}},\ \bibinfo {pages} {4425} (\bibinfo {year}
  {2021})}\BibitemShut {NoStop}%
\bibitem [{\citenamefont {Wu}\ \emph {et~al.}(2023)\citenamefont {Wu},
  \citenamefont {Tao}, \citenamefont {Fan}, \citenamefont {Cui},\ and\
  \citenamefont {Zhang}}]{Wu:2023}%
  \BibitemOpen
  \bibfield  {author} {\bibinfo {author} {\bibfnamefont {Y.}~\bibnamefont
  {Wu}}, \bibinfo {author} {\bibfnamefont {Y.}~\bibnamefont {Tao}}, \bibinfo
  {author} {\bibfnamefont {D.}~\bibnamefont {Fan}}, \bibinfo {author}
  {\bibfnamefont {C.}~\bibnamefont {Cui}},\ and\ \bibinfo {author}
  {\bibfnamefont {Y.}~\bibnamefont {Zhang}},\ }\bibfield  {title} {\bibinfo
  {title} {{Galaxy spectral classification and feature analysis based on
  convolutional neural network}},\ }\href
  {https://doi.org/10.1093/mnras/stad2913} {\bibfield  {journal} {\bibinfo
  {journal} {Monthly Notices of the Royal Astronomical Society}\ }\textbf
  {\bibinfo {volume} {527}},\ \bibinfo {pages} {1163} (\bibinfo {year}
  {2023})}\BibitemShut {NoStop}%
\bibitem [{\citenamefont {Gebhard}\ \emph {et~al.}(2019)\citenamefont
  {Gebhard}, \citenamefont {Kilbertus}, \citenamefont {Harry},\ and\
  \citenamefont {Sch\"olkopf}}]{Gebhard:2019}%
  \BibitemOpen
  \bibfield  {author} {\bibinfo {author} {\bibfnamefont {T.~D.}\ \bibnamefont
  {Gebhard}}, \bibinfo {author} {\bibfnamefont {N.}~\bibnamefont {Kilbertus}},
  \bibinfo {author} {\bibfnamefont {I.}~\bibnamefont {Harry}},\ and\ \bibinfo
  {author} {\bibfnamefont {B.}~\bibnamefont {Sch\"olkopf}},\ }\bibfield
  {title} {\bibinfo {title} {Convolutional neural networks: A magic bullet for
  gravitational-wave detection?},\ }\href
  {https://doi.org/10.1103/PhysRevD.100.063015} {\bibfield  {journal} {\bibinfo
   {journal} {Phys. Rev. D}\ }\textbf {\bibinfo {volume} {100}},\ \bibinfo
  {pages} {063015} (\bibinfo {year} {2019})}\BibitemShut {NoStop}%
\bibitem [{\citenamefont {Li}\ \emph {et~al.}(2020)\citenamefont {Li},
  \citenamefont {Yu}, \citenamefont {Fan},\ and\ \citenamefont
  {Babu}}]{Li:2020}%
  \BibitemOpen
  \bibfield  {author} {\bibinfo {author} {\bibfnamefont {X.-R.}\ \bibnamefont
  {Li}}, \bibinfo {author} {\bibfnamefont {W.-L.}\ \bibnamefont {Yu}}, \bibinfo
  {author} {\bibfnamefont {X.-L.}\ \bibnamefont {Fan}},\ and\ \bibinfo {author}
  {\bibfnamefont {G.~J.}\ \bibnamefont {Babu}},\ }\bibfield  {title} {\bibinfo
  {title} {Some optimizations on detecting gravitational wave using
  convolutional neural network},\ }\bibfield  {journal} {\bibinfo  {journal}
  {Frontiers of Physics}\ }\textbf {\bibinfo {volume} {15}},\ \href
  {https://doi.org/10.1007/s11467-020-0966-4} {10.1007/s11467-020-0966-4}
  (\bibinfo {year} {2020})\BibitemShut {NoStop}%
\bibitem [{\citenamefont {Xia}\ \emph {et~al.}(2021)\citenamefont {Xia},
  \citenamefont {Shao}, \citenamefont {Zhao},\ and\ \citenamefont
  {Cao}}]{Xia:2021}%
  \BibitemOpen
  \bibfield  {author} {\bibinfo {author} {\bibfnamefont {H.}~\bibnamefont
  {Xia}}, \bibinfo {author} {\bibfnamefont {L.}~\bibnamefont {Shao}}, \bibinfo
  {author} {\bibfnamefont {J.}~\bibnamefont {Zhao}},\ and\ \bibinfo {author}
  {\bibfnamefont {Z.}~\bibnamefont {Cao}},\ }\bibfield  {title} {\bibinfo
  {title} {Improved deep learning techniques in gravitational-wave data
  analysis},\ }\href {https://doi.org/10.1103/PhysRevD.103.024040} {\bibfield
  {journal} {\bibinfo  {journal} {Phys. Rev. D}\ }\textbf {\bibinfo {volume}
  {103}},\ \bibinfo {pages} {024040} (\bibinfo {year} {2021})}\BibitemShut
  {NoStop}%
\bibitem [{\citenamefont {Baltus}\ \emph {et~al.}(2021)\citenamefont {Baltus},
  \citenamefont {Janquart}, \citenamefont {Lopez}, \citenamefont {Reza},
  \citenamefont {Caudill},\ and\ \citenamefont {Cudell}}]{Baltus:2021}%
  \BibitemOpen
  \bibfield  {author} {\bibinfo {author} {\bibfnamefont {G.}~\bibnamefont
  {Baltus}}, \bibinfo {author} {\bibfnamefont {J.}~\bibnamefont {Janquart}},
  \bibinfo {author} {\bibfnamefont {M.}~\bibnamefont {Lopez}}, \bibinfo
  {author} {\bibfnamefont {A.}~\bibnamefont {Reza}}, \bibinfo {author}
  {\bibfnamefont {S.}~\bibnamefont {Caudill}},\ and\ \bibinfo {author}
  {\bibfnamefont {J.-R.}\ \bibnamefont {Cudell}},\ }\bibfield  {title}
  {\bibinfo {title} {Convolutional neural networks for the detection of the
  early inspiral of a gravitational-wave signal},\ }\href
  {https://doi.org/10.1103/PhysRevD.103.102003} {\bibfield  {journal} {\bibinfo
   {journal} {Phys. Rev. D}\ }\textbf {\bibinfo {volume} {103}},\ \bibinfo
  {pages} {102003} (\bibinfo {year} {2021})}\BibitemShut {NoStop}%
\bibitem [{\citenamefont {Zhang}\ \emph {et~al.}(2022)\citenamefont {Zhang},
  \citenamefont {Messenger}, \citenamefont {Korsakova}, \citenamefont {Chan},
  \citenamefont {Hu},\ and\ \citenamefont {Zhang}}]{Zhang:2022}%
  \BibitemOpen
  \bibfield  {author} {\bibinfo {author} {\bibfnamefont {X.-T.}\ \bibnamefont
  {Zhang}}, \bibinfo {author} {\bibfnamefont {C.}~\bibnamefont {Messenger}},
  \bibinfo {author} {\bibfnamefont {N.}~\bibnamefont {Korsakova}}, \bibinfo
  {author} {\bibfnamefont {M.~L.}\ \bibnamefont {Chan}}, \bibinfo {author}
  {\bibfnamefont {Y.-M.}\ \bibnamefont {Hu}},\ and\ \bibinfo {author}
  {\bibfnamefont {J.-d.}\ \bibnamefont {Zhang}},\ }\bibfield  {title} {\bibinfo
  {title} {Detecting gravitational waves from extreme mass ratio inspirals
  using convolutional neural networks},\ }\href
  {https://doi.org/10.1103/PhysRevD.105.123027} {\bibfield  {journal} {\bibinfo
   {journal} {Phys. Rev. D}\ }\textbf {\bibinfo {volume} {105}},\ \bibinfo
  {pages} {123027} (\bibinfo {year} {2022})}\BibitemShut {NoStop}%
\bibitem [{\citenamefont {Farsian}\ \emph {et~al.}(2020)\citenamefont
  {Farsian}, \citenamefont {Krachmalnicoff},\ and\ \citenamefont
  {Baccigalupi}}]{Farsian:2020}%
  \BibitemOpen
  \bibfield  {author} {\bibinfo {author} {\bibfnamefont {F.}~\bibnamefont
  {Farsian}}, \bibinfo {author} {\bibfnamefont {N.}~\bibnamefont
  {Krachmalnicoff}},\ and\ \bibinfo {author} {\bibfnamefont {C.}~\bibnamefont
  {Baccigalupi}},\ }\bibfield  {title} {\bibinfo {title} {Foreground model
  recognition through neural networks for cmb b-mode observations},\ }\href
  {https://doi.org/10.1088/1475-7516/2020/07/017} {\bibfield  {journal}
  {\bibinfo  {journal} {Journal of Cosmology and Astroparticle Physics}\
  }\textbf {\bibinfo {volume} {2020}}\bibinfo  {number} { (07)},\ \bibinfo
  {pages} {017–017}}\BibitemShut {NoStop}%
\bibitem [{\citenamefont {{Casas}}\ \emph {et~al.}(2022)\citenamefont
  {{Casas}}, \citenamefont {{Bonavera}}, \citenamefont {{Gonz{\'a}lez-Nuevo}},
  \citenamefont {{Baccigalupi}}, \citenamefont {{Cueli}}, \citenamefont
  {{Crespo}}, \citenamefont {{Goitia}}, \citenamefont {{Santos}}, \citenamefont
  {{S{\'a}nchez}},\ and\ \citenamefont {{de Cos}}}]{Casas:2022}%
  \BibitemOpen
\bibfield  {number} {  }\bibfield  {author} {\bibinfo {author} {\bibfnamefont
  {J.~M.}\ \bibnamefont {{Casas}}}, \bibinfo {author} {\bibfnamefont
  {L.}~\bibnamefont {{Bonavera}}}, \bibinfo {author} {\bibfnamefont
  {J.}~\bibnamefont {{Gonz{\'a}lez-Nuevo}}}, \bibinfo {author} {\bibfnamefont
  {C.}~\bibnamefont {{Baccigalupi}}}, \bibinfo {author} {\bibfnamefont {M.~M.}\
  \bibnamefont {{Cueli}}}, \bibinfo {author} {\bibfnamefont {D.}~\bibnamefont
  {{Crespo}}}, \bibinfo {author} {\bibfnamefont {E.}~\bibnamefont {{Goitia}}},
  \bibinfo {author} {\bibfnamefont {J.~D.}\ \bibnamefont {{Santos}}}, \bibinfo
  {author} {\bibfnamefont {M.~L.}\ \bibnamefont {{S{\'a}nchez}}},\ and\
  \bibinfo {author} {\bibfnamefont {F.~J.}\ \bibnamefont {{de Cos}}},\
  }\bibfield  {title} {\bibinfo {title} {{CENN: A fully convolutional neural
  network for CMB recovery in realistic microwave sky simulations}},\ }\href
  {https://doi.org/10.1051/0004-6361/202243450} {\bibfield  {journal} {\bibinfo
   {journal} {Astronomy \& Astrophysics}\ }\textbf {\bibinfo {volume} {666}},\
  \bibinfo {eid} {A89} (\bibinfo {year} {2022})},\ \Eprint
  {https://arxiv.org/abs/2205.05623} {arXiv:2205.05623 [astro-ph.CO]}
  \BibitemShut {NoStop}%
\bibitem [{\citenamefont {Yan}\ \emph {et~al.}(2024)\citenamefont {Yan},
  \citenamefont {Li}, \citenamefont {Wang}, \citenamefont {Zhang},\ and\
  \citenamefont {Xia}}]{Yan:2024}%
  \BibitemOpen
  \bibfield  {author} {\bibinfo {author} {\bibfnamefont {Y.-P.}\ \bibnamefont
  {Yan}}, \bibinfo {author} {\bibfnamefont {S.-Y.}\ \bibnamefont {Li}},
  \bibinfo {author} {\bibfnamefont {G.-J.}\ \bibnamefont {Wang}}, \bibinfo
  {author} {\bibfnamefont {Z.}~\bibnamefont {Zhang}},\ and\ \bibinfo {author}
  {\bibfnamefont {J.-Q.}\ \bibnamefont {Xia}},\ }\href
  {https://arxiv.org/abs/2406.17685} {\bibinfo {title} {Cmbfscnn: Cosmic
  microwave background polarization foreground subtraction with convolutional
  neural network}} (\bibinfo {year} {2024}),\ \Eprint
  {https://arxiv.org/abs/2406.17685} {arXiv:2406.17685 [astro-ph.CO]}
  \BibitemShut {NoStop}%
\bibitem [{\citenamefont {Caron}\ \emph {et~al.}(2018)\citenamefont {Caron},
  \citenamefont {Gómez-Vargas}, \citenamefont {Hendriks},\ and\ \citenamefont
  {de~Austri}}]{Caron:2018}%
  \BibitemOpen
  \bibfield  {author} {\bibinfo {author} {\bibfnamefont {S.}~\bibnamefont
  {Caron}}, \bibinfo {author} {\bibfnamefont {G.~A.}\ \bibnamefont
  {Gómez-Vargas}}, \bibinfo {author} {\bibfnamefont {L.}~\bibnamefont
  {Hendriks}},\ and\ \bibinfo {author} {\bibfnamefont {R.~R.}\ \bibnamefont
  {de~Austri}},\ }\bibfield  {title} {\bibinfo {title} {Analyzing $\gamma$ rays
  of the galactic center with deep learning},\ }\href
  {https://doi.org/10.1088/1475-7516/2018/05/058} {\bibfield  {journal}
  {\bibinfo  {journal} {Journal of Cosmology and Astroparticle Physics}\
  }\textbf {\bibinfo {volume} {2018}}\bibinfo  {number} { (05)},\ \bibinfo
  {pages} {058}}\BibitemShut {NoStop}%
\bibitem [{\citenamefont {Khek}\ \emph {et~al.}(2022)\citenamefont {Khek},
  \citenamefont {Mishra}, \citenamefont {Buuck},\ and\ \citenamefont
  {Shutt}}]{Khek:2022}%
  \BibitemOpen
\bibfield  {number} {  }\bibfield  {author} {\bibinfo {author} {\bibfnamefont
  {B.}~\bibnamefont {Khek}}, \bibinfo {author} {\bibfnamefont {A.}~\bibnamefont
  {Mishra}}, \bibinfo {author} {\bibfnamefont {M.}~\bibnamefont {Buuck}},\ and\
  \bibinfo {author} {\bibfnamefont {T.}~\bibnamefont {Shutt}},\ }\bibfield
  {title} {\bibinfo {title} {Gamma ray source localization for time projection
  chamber telescopes using convolutional neural networks},\ }\href
  {https://doi.org/10.3390/ai3040058} {\bibfield  {journal} {\bibinfo
  {journal} {AI}\ }\textbf {\bibinfo {volume} {3}},\ \bibinfo {pages} {975}
  (\bibinfo {year} {2022})}\BibitemShut {NoStop}%
\bibitem [{\citenamefont {Khosa}\ \emph {et~al.}(2020)\citenamefont {Khosa},
  \citenamefont {Mars}, \citenamefont {Richards},\ and\ \citenamefont
  {Sanz}}]{Khosa:2020}%
  \BibitemOpen
  \bibfield  {author} {\bibinfo {author} {\bibfnamefont {C.~K.}\ \bibnamefont
  {Khosa}}, \bibinfo {author} {\bibfnamefont {L.}~\bibnamefont {Mars}},
  \bibinfo {author} {\bibfnamefont {J.}~\bibnamefont {Richards}},\ and\
  \bibinfo {author} {\bibfnamefont {V.}~\bibnamefont {Sanz}},\ }\bibfield
  {title} {\bibinfo {title} {Convolutional neural networks for direct detection
  of dark matter},\ }\href {https://doi.org/10.1088/1361-6471/ab8e94}
  {\bibfield  {journal} {\bibinfo  {journal} {Journal of Physics G: Nuclear and
  Particle Physics}\ }\textbf {\bibinfo {volume} {47}},\ \bibinfo {pages}
  {095201} (\bibinfo {year} {2020})}\BibitemShut {NoStop}%
\bibitem [{\citenamefont {Pritchard}\ and\ \citenamefont
  {Loeb}(2012)}]{Pritchard_2012}%
  \BibitemOpen
  \bibfield  {author} {\bibinfo {author} {\bibfnamefont {J.~R.}\ \bibnamefont
  {Pritchard}}\ and\ \bibinfo {author} {\bibfnamefont {A.}~\bibnamefont
  {Loeb}},\ }\bibfield  {title} {\bibinfo {title} {21 cm cosmology in the 21st
  century},\ }\href {https://doi.org/10.1088/0034-4885/75/8/086901} {\bibfield
  {journal} {\bibinfo  {journal} {Rep. Prog. Phys}\ }\textbf {\bibinfo {volume}
  {75}},\ \bibinfo {pages} {086901} (\bibinfo {year} {2012})}\BibitemShut
  {NoStop}%
\bibitem [{\citenamefont {Natwariya}(2023)}]{Natwariya:2023T}%
  \BibitemOpen
  \bibfield  {author} {\bibinfo {author} {\bibfnamefont {P.~K.}\ \bibnamefont
  {Natwariya}},\ }\href@noop {} {\bibinfo {title} {21 cm line astronomy and
  constraining new physics}} (\bibinfo {year} {2023}),\ \Eprint
  {https://arxiv.org/abs/2301.02655} {arXiv:2301.02655 [astro-ph.CO]}
  \BibitemShut {NoStop}%
\bibitem [{\citenamefont {{Wouthuysen}}(1952)}]{1952AJ.....57R..31W}%
  \BibitemOpen
  \bibfield  {author} {\bibinfo {author} {\bibfnamefont {S.~A.}\ \bibnamefont
  {{Wouthuysen}}},\ }\bibfield  {title} {\bibinfo {title} {{On the excitation
  mechanism of the 21-cm (radio-frequency) interstellar hydrogen emission
  line.}},\ }\href {https://doi.org/10.1086/106661} {\bibfield  {journal}
  {\bibinfo  {journal} {The Astronomical Journal}\ }\textbf {\bibinfo {volume}
  {57}},\ \bibinfo {pages} {31} (\bibinfo {year} {1952})}\BibitemShut {NoStop}%
\bibitem [{\citenamefont {{Field}}(1958{\natexlab{a}})}]{Field}%
  \BibitemOpen
  \bibfield  {author} {\bibinfo {author} {\bibfnamefont {G.~B.}\ \bibnamefont
  {{Field}}},\ }\bibfield  {title} {\bibinfo {title} {Excitation of the
  hydrogen 21-cm line},\ }\href {https://doi.org/10.1109/JRPROC.1958.286741}
  {\bibfield  {journal} {\bibinfo  {journal} {Proceedings of the IRE}\ }\textbf
  {\bibinfo {volume} {46}},\ \bibinfo {pages} {240} (\bibinfo {year}
  {1958}{\natexlab{a}})}\BibitemShut {NoStop}%
\bibitem [{\citenamefont {{Field}}(1958{\natexlab{b}})}]{1958PIRE...46..240F}%
  \BibitemOpen
  \bibfield  {author} {\bibinfo {author} {\bibfnamefont {G.~B.}\ \bibnamefont
  {{Field}}},\ }\bibfield  {title} {\bibinfo {title} {{Excitation of the
  Hydrogen 21-CM Line}},\ }\href {https://doi.org/10.1109/JRPROC.1958.286741}
  {\bibfield  {journal} {\bibinfo  {journal} {Proceedings of the IRE}\ }\textbf
  {\bibinfo {volume} {46}},\ \bibinfo {pages} {240} (\bibinfo {year}
  {1958}{\natexlab{b}})}\BibitemShut {NoStop}%
\bibitem [{\citenamefont {{Hirata}}(2006)}]{2006MNRAS.367..259H}%
  \BibitemOpen
  \bibfield  {author} {\bibinfo {author} {\bibfnamefont {C.~M.}\ \bibnamefont
  {{Hirata}}},\ }\bibfield  {title} {\bibinfo {title} {{Wouthuysen-Field
  coupling strength and application to high-redshift 21-cm radiation}},\ }\href
  {https://doi.org/10.1111/j.1365-2966.2005.09949.x} {\bibfield  {journal}
  {\bibinfo  {journal} {\mnras}\ }\textbf {\bibinfo {volume} {367}},\ \bibinfo
  {pages} {259} (\bibinfo {year} {2006})},\ \Eprint
  {https://arxiv.org/abs/astro-ph/0507102} {arXiv:astro-ph/0507102 [astro-ph]}
  \BibitemShut {NoStop}%
\bibitem [{\citenamefont {{Mesinger}}\ \emph {et~al.}(2011)\citenamefont
  {{Mesinger}}, \citenamefont {{Furlanetto}},\ and\ \citenamefont
  {{Cen}}}]{2011MNRAS.411..955M}%
  \BibitemOpen
  \bibfield  {author} {\bibinfo {author} {\bibfnamefont {A.}~\bibnamefont
  {{Mesinger}}}, \bibinfo {author} {\bibfnamefont {S.}~\bibnamefont
  {{Furlanetto}}},\ and\ \bibinfo {author} {\bibfnamefont {R.}~\bibnamefont
  {{Cen}}},\ }\bibfield  {title} {\bibinfo {title} {{21CMFAST: a fast,
  seminumerical simulation of the high-redshift 21-cm signal}},\ }\href
  {https://doi.org/10.1111/j.1365-2966.2010.17731.x} {\bibfield  {journal}
  {\bibinfo  {journal} {\mnras}\ }\textbf {\bibinfo {volume} {411}},\ \bibinfo
  {pages} {955} (\bibinfo {year} {2011})},\ \Eprint
  {https://arxiv.org/abs/1003.3878} {arXiv:1003.3878 [astro-ph.CO]}
  \BibitemShut {NoStop}%
\bibitem [{\citenamefont {{Pritchard}}\ and\ \citenamefont
  {{Loeb}}(2012)}]{2012RPPh...75h6901P}%
  \BibitemOpen
  \bibfield  {author} {\bibinfo {author} {\bibfnamefont {J.~R.}\ \bibnamefont
  {{Pritchard}}}\ and\ \bibinfo {author} {\bibfnamefont {A.}~\bibnamefont
  {{Loeb}}},\ }\bibfield  {title} {\bibinfo {title} {{21 cm cosmology in the
  21st century}},\ }\href {https://doi.org/10.1088/0034-4885/75/8/086901}
  {\bibfield  {journal} {\bibinfo  {journal} {Reports on Progress in Physics}\
  }\textbf {\bibinfo {volume} {75}},\ \bibinfo {eid} {086901} (\bibinfo {year}
  {2012})},\ \Eprint {https://arxiv.org/abs/1109.6012} {arXiv:1109.6012
  [astro-ph.CO]} \BibitemShut {NoStop}%
\bibitem [{\citenamefont {{Seager}}\ \emph {et~al.}(1999)\citenamefont
  {{Seager}}, \citenamefont {{Sasselov}},\ and\ \citenamefont
  {{Scott}}}]{1999ApJ...523L...1S}%
  \BibitemOpen
  \bibfield  {author} {\bibinfo {author} {\bibfnamefont {S.}~\bibnamefont
  {{Seager}}}, \bibinfo {author} {\bibfnamefont {D.~D.}\ \bibnamefont
  {{Sasselov}}},\ and\ \bibinfo {author} {\bibfnamefont {D.}~\bibnamefont
  {{Scott}}},\ }\bibfield  {title} {\bibinfo {title} {{A New Calculation of the
  Recombination Epoch}},\ }\href {https://doi.org/10.1086/312250} {\bibfield
  {journal} {\bibinfo  {journal} {\apjl}\ }\textbf {\bibinfo {volume} {523}},\
  \bibinfo {pages} {L1} (\bibinfo {year} {1999})},\ \Eprint
  {https://arxiv.org/abs/astro-ph/9909275} {arXiv:astro-ph/9909275 [astro-ph]}
  \BibitemShut {NoStop}%
\bibitem [{\citenamefont {{Seager}}\ \emph {et~al.}(2000)\citenamefont
  {{Seager}}, \citenamefont {{Sasselov}},\ and\ \citenamefont
  {{Scott}}}]{2000ApJS..128..407S}%
  \BibitemOpen
  \bibfield  {author} {\bibinfo {author} {\bibfnamefont {S.}~\bibnamefont
  {{Seager}}}, \bibinfo {author} {\bibfnamefont {D.~D.}\ \bibnamefont
  {{Sasselov}}},\ and\ \bibinfo {author} {\bibfnamefont {D.}~\bibnamefont
  {{Scott}}},\ }\bibfield  {title} {\bibinfo {title} {{How Exactly Did the
  Universe Become Neutral?}},\ }\href {https://doi.org/10.1086/313388}
  {\bibfield  {journal} {\bibinfo  {journal} {\apjs}\ }\textbf {\bibinfo
  {volume} {128}},\ \bibinfo {pages} {407} (\bibinfo {year} {2000})},\ \Eprint
  {https://arxiv.org/abs/astro-ph/9912182} {arXiv:astro-ph/9912182 [astro-ph]}
  \BibitemShut {NoStop}%
\bibitem [{\citenamefont {{Contino}}\ \emph {et~al.}(2019)\citenamefont
  {{Contino}}, \citenamefont {{Mitridate}}, \citenamefont {{Podo}},\ and\
  \citenamefont {{Redi}}}]{2019JHEP...02..187C}%
  \BibitemOpen
  \bibfield  {author} {\bibinfo {author} {\bibfnamefont {R.}~\bibnamefont
  {{Contino}}}, \bibinfo {author} {\bibfnamefont {A.}~\bibnamefont
  {{Mitridate}}}, \bibinfo {author} {\bibfnamefont {A.}~\bibnamefont
  {{Podo}}},\ and\ \bibinfo {author} {\bibfnamefont {M.}~\bibnamefont
  {{Redi}}},\ }\bibfield  {title} {\bibinfo {title} {{Gluequark dark matter}},\
  }\href {https://doi.org/10.1007/JHEP02(2019)187} {\bibfield  {journal}
  {\bibinfo  {journal} {Journal of High Energy Physics}\ }\textbf {\bibinfo
  {volume} {2019}},\ \bibinfo {eid} {187} (\bibinfo {year} {2019})},\ \Eprint
  {https://arxiv.org/abs/1811.06975} {arXiv:1811.06975 [hep-ph]} \BibitemShut
  {NoStop}%
\bibitem [{\citenamefont {{D'Amico}}\ \emph {et~al.}(2018)\citenamefont
  {{D'Amico}}, \citenamefont {{Panci}},\ and\ \citenamefont
  {{Strumia}}}]{2018PhRvL.121a1103D}%
  \BibitemOpen
  \bibfield  {author} {\bibinfo {author} {\bibfnamefont {G.}~\bibnamefont
  {{D'Amico}}}, \bibinfo {author} {\bibfnamefont {P.}~\bibnamefont {{Panci}}},\
  and\ \bibinfo {author} {\bibfnamefont {A.}~\bibnamefont {{Strumia}}},\
  }\bibfield  {title} {\bibinfo {title} {{Bounds on Dark-Matter Annihilations
  from 21-cm Data}},\ }\href {https://doi.org/10.1103/PhysRevLett.121.011103}
  {\bibfield  {journal} {\bibinfo  {journal} {\prl}\ }\textbf {\bibinfo
  {volume} {121}},\ \bibinfo {eid} {011103} (\bibinfo {year} {2018})},\ \Eprint
  {https://arxiv.org/abs/1803.03629} {arXiv:1803.03629 [astro-ph.CO]}
  \BibitemShut {NoStop}%
\bibitem [{\citenamefont {{Galli}}\ \emph {et~al.}(2009)\citenamefont
  {{Galli}}, \citenamefont {{Iocco}}, \citenamefont {{Bertone}},\ and\
  \citenamefont {{Melchiorri}}}]{2009PhRvD..80b3505G}%
  \BibitemOpen
  \bibfield  {author} {\bibinfo {author} {\bibfnamefont {S.}~\bibnamefont
  {{Galli}}}, \bibinfo {author} {\bibfnamefont {F.}~\bibnamefont {{Iocco}}},
  \bibinfo {author} {\bibfnamefont {G.}~\bibnamefont {{Bertone}}},\ and\
  \bibinfo {author} {\bibfnamefont {A.}~\bibnamefont {{Melchiorri}}},\
  }\bibfield  {title} {\bibinfo {title} {{CMB constraints on dark matter models
  with large annihilation cross section}},\ }\href
  {https://doi.org/10.1103/PhysRevD.80.023505} {\bibfield  {journal} {\bibinfo
  {journal} {\prd}\ }\textbf {\bibinfo {volume} {80}},\ \bibinfo {eid} {023505}
  (\bibinfo {year} {2009})},\ \Eprint {https://arxiv.org/abs/0905.0003}
  {arXiv:0905.0003 [astro-ph.CO]} \BibitemShut {NoStop}%
\bibitem [{\citenamefont {{Ali-Ha{\"\i}moud}}\ and\ \citenamefont
  {{Hirata}}(2011)}]{2011PhRvD..83d3513A}%
  \BibitemOpen
  \bibfield  {author} {\bibinfo {author} {\bibfnamefont {Y.}~\bibnamefont
  {{Ali-Ha{\"\i}moud}}}\ and\ \bibinfo {author} {\bibfnamefont {C.~M.}\
  \bibnamefont {{Hirata}}},\ }\bibfield  {title} {\bibinfo {title} {{HyRec: A
  fast and highly accurate primordial hydrogen and helium recombination
  code}},\ }\href {https://doi.org/10.1103/PhysRevD.83.043513} {\bibfield
  {journal} {\bibinfo  {journal} {\prd}\ }\textbf {\bibinfo {volume} {83}},\
  \bibinfo {eid} {043513} (\bibinfo {year} {2011})},\ \Eprint
  {https://arxiv.org/abs/1011.3758} {arXiv:1011.3758 [astro-ph.CO]}
  \BibitemShut {NoStop}%
\bibitem [{\citenamefont {Peebles}(1968)}]{Peebles:1968ja}%
  \BibitemOpen
  \bibfield  {author} {\bibinfo {author} {\bibfnamefont {P.~J.~E.}\
  \bibnamefont {Peebles}},\ }\bibfield  {title} {\bibinfo {title}
  {{Recombination of the Primeval Plasma}},\ }\href
  {https://doi.org/10.1086/149628} {\bibfield  {journal} {\bibinfo  {journal}
  {Astrophys. J.}\ }\textbf {\bibinfo {volume} {153}},\ \bibinfo {pages} {1}
  (\bibinfo {year} {1968})}\BibitemShut {NoStop}%
\bibitem [{\citenamefont {Ali-Haimoud}\ and\ \citenamefont
  {Hirata}(2011)}]{AliHaimoud:2010dx}%
  \BibitemOpen
  \bibfield  {author} {\bibinfo {author} {\bibfnamefont {Y.}~\bibnamefont
  {Ali-Haimoud}}\ and\ \bibinfo {author} {\bibfnamefont {C.~M.}\ \bibnamefont
  {Hirata}},\ }\bibfield  {title} {\bibinfo {title} {{HyRec: A fast and highly
  accurate primordial hydrogen and helium recombination code}},\ }\href
  {https://doi.org/10.1103/PhysRevD.83.043513} {\bibfield  {journal} {\bibinfo
  {journal} {Phys. Rev.}\ }\textbf {\bibinfo {volume} {D83}},\ \bibinfo {pages}
  {043513} (\bibinfo {year} {2011})}\BibitemShut {NoStop}%
\bibitem [{\citenamefont {{Tung}}\ and\ \citenamefont
  {{Chan}}(1984)}]{1984PRA...30....1175P}%
  \BibitemOpen
  \bibfield  {author} {\bibinfo {author} {\bibfnamefont {S.~X.~M.}\
  \bibnamefont {{Tung}}, \bibfnamefont {J.~H.}}\ and\ \bibinfo {author}
  {\bibfnamefont {F.~T.}\ \bibnamefont {{Chan}}},\ }\bibfield  {title}
  {\bibinfo {title} {{Two-photon decay of hydrogenic atoms}},\ }\href
  {https://doi.org/10.1103/PhysRevA.30.1175} {\bibfield  {journal} {\bibinfo
  {journal} {\pra}\ }\textbf {\bibinfo {volume} {30}},\ \bibinfo {pages} {1175}
  (\bibinfo {year} {1984})}\BibitemShut {NoStop}%
\bibitem [{\citenamefont {{Holland}}(1992)}]{GA1992}%
  \BibitemOpen
  \bibfield  {author} {\bibinfo {author} {\bibfnamefont {J.~H.}\ \bibnamefont
  {{Holland}}},\ }\bibfield  {title} {\bibinfo {title} {{Genetic Algorithms}},\
  }\href@noop {} {\bibfield  {journal} {\bibinfo  {journal} {Scientific
  American}\ }\textbf {\bibinfo {volume} {267}},\ \bibinfo {pages} {66}
  (\bibinfo {year} {1992})}\BibitemShut {NoStop}%
\bibitem [{\citenamefont {Cirelli}\ \emph {et~al.}(2023)\citenamefont
  {Cirelli}, \citenamefont {Fornengo}, \citenamefont {Koechler}, \citenamefont
  {Pinetti},\ and\ \citenamefont {Roach}}]{Cirelli:2023tnx}%
  \BibitemOpen
  \bibfield  {author} {\bibinfo {author} {\bibfnamefont {M.}~\bibnamefont
  {Cirelli}}, \bibinfo {author} {\bibfnamefont {N.}~\bibnamefont {Fornengo}},
  \bibinfo {author} {\bibfnamefont {J.}~\bibnamefont {Koechler}}, \bibinfo
  {author} {\bibfnamefont {E.}~\bibnamefont {Pinetti}},\ and\ \bibinfo {author}
  {\bibfnamefont {B.~M.}\ \bibnamefont {Roach}},\ }\bibfield  {title} {\bibinfo
  {title} {{Putting all the X in one basket: Updated X-ray constraints on
  sub-GeV Dark Matter}},\ }\href
  {https://doi.org/10.1088/1475-7516/2023/07/026} {\bibfield  {journal}
  {\bibinfo  {journal} {JCAP}\ }\textbf {\bibinfo {volume} {07}},\ \bibinfo
  {pages} {026}},\ \Eprint {https://arxiv.org/abs/2303.08854} {arXiv:2303.08854
  [hep-ph]} \BibitemShut {NoStop}%
\bibitem [{\citenamefont {{Slatyer}}(2016)}]{2016PhRvD..93b3527S}%
  \BibitemOpen
  \bibfield  {author} {\bibinfo {author} {\bibfnamefont {T.~R.}\ \bibnamefont
  {{Slatyer}}},\ }\bibfield  {title} {\bibinfo {title} {{Indirect dark matter
  signatures in the cosmic dark ages. I. Generalizing the bound on s -wave dark
  matter annihilation from Planck results}},\ }\href
  {https://doi.org/10.1103/PhysRevD.93.023527} {\bibfield  {journal} {\bibinfo
  {journal} {\prd}\ }\textbf {\bibinfo {volume} {93}},\ \bibinfo {eid} {023527}
  (\bibinfo {year} {2016})},\ \Eprint {https://arxiv.org/abs/1506.03811}
  {arXiv:1506.03811 [hep-ph]} \BibitemShut {NoStop}%
\bibitem [{\citenamefont {Evoli}\ \emph {et~al.}(2012)\citenamefont {Evoli},
  \citenamefont {Valdés}, \citenamefont {Ferrara},\ and\ \citenamefont
  {Yoshida}}]{Evoli:2012}%
  \BibitemOpen
  \bibfield  {author} {\bibinfo {author} {\bibfnamefont {C.}~\bibnamefont
  {Evoli}}, \bibinfo {author} {\bibfnamefont {M.}~\bibnamefont {Valdés}},
  \bibinfo {author} {\bibfnamefont {A.}~\bibnamefont {Ferrara}},\ and\ \bibinfo
  {author} {\bibfnamefont {N.}~\bibnamefont {Yoshida}},\ }\bibfield  {title}
  {\bibinfo {title} {Energy deposition by weakly interacting massive particles:
  a comprehensive study},\ }\href
  {https://doi.org/10.1111/j.1365-2966.2012.20624.x} {\bibfield  {journal}
  {\bibinfo  {journal} {Monthly Notices of the Royal Astronomical Society}\
  }\textbf {\bibinfo {volume} {422}},\ \bibinfo {pages} {420} (\bibinfo {year}
  {2012})}\BibitemShut {NoStop}%
\bibitem [{\citenamefont {Slatyer}\ \emph {et~al.}(2009)\citenamefont
  {Slatyer}, \citenamefont {Padmanabhan},\ and\ \citenamefont
  {Finkbeiner}}]{Slatyer:2009}%
  \BibitemOpen
  \bibfield  {author} {\bibinfo {author} {\bibfnamefont {T.~R.}\ \bibnamefont
  {Slatyer}}, \bibinfo {author} {\bibfnamefont {N.}~\bibnamefont
  {Padmanabhan}},\ and\ \bibinfo {author} {\bibfnamefont {D.~P.}\ \bibnamefont
  {Finkbeiner}},\ }\bibfield  {title} {\bibinfo {title} {Cmb constraints on
  wimp annihilation: Energy absorption during the recombination epoch},\ }\href
  {https://doi.org/10.1103/PhysRevD.80.043526} {\bibfield  {journal} {\bibinfo
  {journal} {Phys. Rev. D}\ }\textbf {\bibinfo {volume} {80}},\ \bibinfo
  {pages} {043526} (\bibinfo {year} {2009})}\BibitemShut {NoStop}%
\bibitem [{\citenamefont {Valdés}\ \emph {et~al.}(2010)\citenamefont
  {Valdés}, \citenamefont {Evoli},\ and\ \citenamefont
  {Ferrara}}]{Valds:2010}%
  \BibitemOpen
  \bibfield  {author} {\bibinfo {author} {\bibfnamefont {M.}~\bibnamefont
  {Valdés}}, \bibinfo {author} {\bibfnamefont {C.}~\bibnamefont {Evoli}},\
  and\ \bibinfo {author} {\bibfnamefont {A.}~\bibnamefont {Ferrara}},\
  }\bibfield  {title} {\bibinfo {title} {Particle energy cascade in the
  intergalactic medium},\ }\bibfield  {journal} {\bibinfo  {journal} {Monthly
  Notices of the Royal Astronomical Society}\ }\href
  {https://doi.org/10.1111/j.1365-2966.2010.16387.x}
  {10.1111/j.1365-2966.2010.16387.x} (\bibinfo {year} {2010})\BibitemShut
  {NoStop}%
\bibitem [{\citenamefont {{Pourtsidou}}\ \emph {et~al.}(2016)\citenamefont
  {{Pourtsidou}}, \citenamefont {{Bacon}}, \citenamefont {{Crittenden}},\ and\
  \citenamefont {{Metcalf}}}]{2016MNRAS.459..863P}%
  \BibitemOpen
  \bibfield  {author} {\bibinfo {author} {\bibfnamefont {A.}~\bibnamefont
  {{Pourtsidou}}}, \bibinfo {author} {\bibfnamefont {D.}~\bibnamefont
  {{Bacon}}}, \bibinfo {author} {\bibfnamefont {R.}~\bibnamefont
  {{Crittenden}}},\ and\ \bibinfo {author} {\bibfnamefont {R.~B.}\ \bibnamefont
  {{Metcalf}}},\ }\bibfield  {title} {\bibinfo {title} {{Prospects for
  clustering and lensing measurements with forthcoming intensity mapping and
  optical surveys}},\ }\href {https://doi.org/10.1093/mnras/stw658} {\bibfield
  {journal} {\bibinfo  {journal} {\mnras}\ }\textbf {\bibinfo {volume} {459}},\
  \bibinfo {pages} {863} (\bibinfo {year} {2016})},\ \Eprint
  {https://arxiv.org/abs/1509.03286} {arXiv:1509.03286 [astro-ph.CO]}
  \BibitemShut {NoStop}%
\bibitem [{\citenamefont {Sun}\ \emph {et~al.}(2023)\citenamefont {Sun},
  \citenamefont {Foster}, \citenamefont {Liu}, \citenamefont {Muñoz},\ and\
  \citenamefont {Slatyer}}]{Sun:2023}%
  \BibitemOpen
  \bibfield  {author} {\bibinfo {author} {\bibfnamefont {Y.}~\bibnamefont
  {Sun}}, \bibinfo {author} {\bibfnamefont {J.~W.}\ \bibnamefont {Foster}},
  \bibinfo {author} {\bibfnamefont {H.}~\bibnamefont {Liu}}, \bibinfo {author}
  {\bibfnamefont {J.~B.}\ \bibnamefont {Muñoz}},\ and\ \bibinfo {author}
  {\bibfnamefont {T.~R.}\ \bibnamefont {Slatyer}},\ }\href
  {https://arxiv.org/abs/2312.11608} {\bibinfo {title} {Inhomogeneous energy
  injection in the 21-cm power spectrum: Sensitivity to dark matter decay}}
  (\bibinfo {year} {2023}),\ \Eprint {https://arxiv.org/abs/2312.11608}
  {arXiv:2312.11608 [hep-ph]} \BibitemShut {NoStop}%
\end{thebibliography}%

\end{document}